\documentclass[graybox, secnum]{svmult}

\usepackage{mathptmx}       
\usepackage{helvet}         
\usepackage{courier}        
\usepackage{type1cm}        
\usepackage{makeidx}         
\usepackage{graphicx}        
\usepackage{multicol}        
\usepackage[bottom]{footmisc}
\usepackage{hyperref}        
\usepackage{soul}            
\hypersetup{colorlinks=true,urlcolor=blue}
\usepackage[square,numbers]{natbib}
\makeindex             

\begin{document}
\title*{Chapter ``Black Hole Thermodynamics and Perturbative Quantum Gravity''}
\author{Dmitri V. Fursaev}
\institute{Dmitri V. Fursaev \at Dubna State University, Universitetskaya str. 19, Dubna, Moscow Region, Russia, \email{fursaev@theor.jinr.ru}
}
%
%
\maketitle
\abstract{An introduction to generalized thermodynamics of quantum black holes, in the one-loop approximation, is given. The material is aimed at graduate students. The topics include: quantum evaporation of black holes, Euclidean formulation of quantum theory on black hole backgrounds, the Hartle-Hawking-Israel state, generalized entropy of a quantum black hole and its relation to the entropy of entanglement.}

\section*{Keywords} 
Black holes, thermodynamics, statistical mechanics, quantum entanglement.

\section{Introduction}

Although the perturbative quantum gravity approach has a limited range of applicability, its use in the last decades
led to some conceptual issues which are to be addressed in the full-fledged quantum gravity theory.   The most important issues include understanding evaporation of quantum black holes and resolution of the information loss paradox, as well as finding a microscopic origin of black hole entropy.
 
Black holes are specific solutions of the Einstein equations,
\begin{equation}\label{i.1}
R_{\mu\nu}-\frac 12 g_{\mu\nu}R={8\pi G \over c^4}T_{\mu\nu}~~,
\end{equation}
which
describe regions of a space-time where the gravitational field is
so strong that nothing, including light signals, can escape them.
The interior of a black hole is hidden from an external observer.
The boundary of the unobservable region is called the horizon. In (\ref{i.1}) we use standard notations $R_{\mu\nu}$, $R$, $T_{\mu\nu}$
for the Ricci tensor, the scalar curvature, and the stress-energy tensor of matter, respectively. $G$ is the Newton constant. The Schwarzschild and Kerr black holes are solutions to the vacuum equations (\ref{i.1}) with $T_{\mu\nu}=0$. 

In recent years our understanding of physics near the black hole horizon received important experimental evidences on the base of direct detection of gravitational waves from binary black hole mergers \cite{LIGOScientific:2016aoc}
and observations of shadows of the super-massive black holes \cite{EventHorizonTelescope:2022xnr}.

The perturbative quantum gravity, in this Chapter, is treated in the one-loop approximation or as a theory of free quantum fields 
on black hole geometries.   By explaining quantum effects near black holes in these rather restricted models 
we come to important insights which have been a matter of intensive discussions in a large number of publications.

The concrete aim of this Chapter is to give a self-consistent introduction to generalized thermodynamics of quantum black holes, accessible 
to graduate students.  The material is organized as follows. We start in Sec. \ref{def} with a brief description of black hole solutions by focusing mostly on 
the Killing structure of the black hole horizon and near-horizon features which are needed do define the first law of black hole mechanics.
Quantization of free fields on external backgrounds  is presented in Sec. \ref{fields}. The essence of the Hawking effect is discussed in Sec. 
\ref{evap}, by using the so called s-mode approximation. Thermodynamics of classical black holes is discussed in Sec. 
\ref{noether}.  The basic concept, the Hartle-Hawking-Israel state, which we use to study quantum black holes, is introduced in Sec. \ref{QBH}.  We also give here some elements of a spectral theory of second order elliptic operators and define the Euclidean effective action.
From a point of view of stationary observers quantum matter near black hole horizon is in a high-temperature regime. Hence some features of high-temperature hydrodynamics in gravitational fields are discussed in Sec. \ref{RHD}. 
Finally, in Sec. \ref{GTD} we consider generalized thermodynamics of quantum black holes, and, in particular, generalized black hole entropy.
Quantum corrections to the entropy are discussed in detail. We introduce the notion of entanglement entropy
and show that the generalized entropy is partly related to entanglement of 
states across the black hole horizon. Section \ref{further} contains concluding comments .

We include in this Chapter almost all required definitions and try to show how basic relations can be derived . 
We use
the system of units where $\hbar=c=k_B=1$ ($k_B$
is the Boltzmann constant), 
the Lorentzian signature is defined as $(-,+,+,+)$, geometrical conventions coincide 
with \cite{Misner:1973prb}. 

\section{Necessary definitions}\label{def}

We start with a brief description of basic properties of black hole geometries in the near-horizon approximation. For a comprehensive introduction to black hole physics see \cite{Misner:1973prb},\cite{Frolov:1998wf}. 
A metric of a neutral rotating black hole, which is most interesting from the point of view of physical applications, is the Kerr solution to the Einstein equations (\ref{i.1}) in vacuum, $T_{\mu\nu}=0$,
\begin{equation}\label{2.1}
ds^2=g_{tt}dt^2+g_{rr} dr^2+2g_{t\varphi}dt d \varphi+g_{\varphi\varphi}d\varphi^2+g_{\theta\theta} d\theta^2~~,
\end{equation}
\begin{equation}\label{2.2}
g_{tt}=-\left(1-{2MGr \over \Sigma}\right)~~,~~g_{\theta\theta}=\Sigma~~,~~g_{rr}={\Sigma \over \Delta}~~,
\end{equation}
\begin{equation}\label{2.3}
g_{t\varphi}=-{2MGra \over \Sigma}\sin^2\theta~~,~~g_{\varphi\varphi}=((r^2+a^2)^2-a^2\sin^2\theta\Delta){\sin^2\theta \over \Sigma}~~,
\end{equation}
\begin{equation}\label{2.4}
\Sigma=r^2+a^2\cos^2\theta~~,~~\Delta=r^2-2MGr+a^2~~.
\end{equation}
Metric (\ref{2.1}) is written in the Boyer–Lindquist coordinates. The Kerr solution is asymptotically flat at large $r$. By analyzing its 
behavior at large $r$ one concludes that $M$ is the mass of the source, $J=Ma$ is its angular momentum. It is supposed that $MG>a$.

We also need the Schwarzschild solution, which follows from (\ref{2.1})-(\ref{2.4}) when $a=0$,
\begin{equation}\label{2.5}
ds^2=g_{tt}dt^2+g_{rr} dr^2+r^2(\sin^2\theta d\varphi^2+ d\theta^2)~~,
\end{equation}
\begin{equation}\label{2.6}
-g_{tt}=g_{rr}^{-1}=1-{2MG \over r}~~.
\end{equation}

\begin{figure}
		\includegraphics[width=11cm]{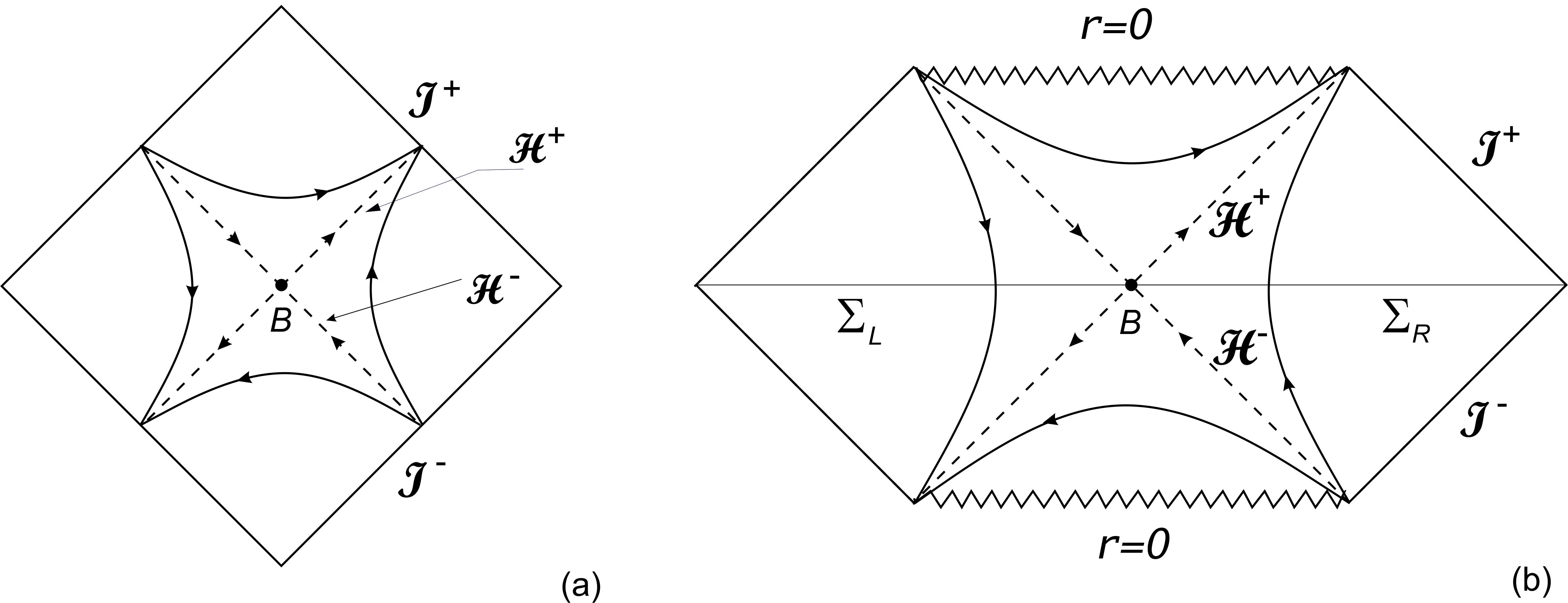} 
		\caption{Carter-Penrose diagrams for Minkowsky space-time ($a$) and for eternal Schwarzschild black hole ($b$). Lines with arrows are integral lines
		of the Killing-vector field. The Killing field in the Minkowsky space-time corresponds to Lorentz boosts. The Killing horizons ${\cal H}^\pm$ intersect at bifurcation 
		2-surfaces. A constant time section $\Sigma_L\cup \Sigma_R$ on the black hole diagram is the Einstein-Rosen bridge.}
		\label{fig2}
\end{figure}

We denote by $\cal H$ the event horizon of a black hole. $\cal H$ as a null hypersurface located at a constant radial coordinate $r$. 
By the definition, the normal vector $l_\mu$ to a null hypersurface is null, $l^2=0$. For constant $r$ hypersurfaces  $l_\mu=\delta_\mu^r$.
Hence a surface $r=r_0$ is null if $g^{rr}(r_0)=0$, or 
$\Delta(r_0)=0$.  This equation has two roots and  $\cal H$ corresponds to the largest root, $r_0=r_H=MG+\sqrt{(MG)^2-a^2}$.
For eternal black holes (see Fig. (\ref{fig2})) the horizon $\cal H={\cal H}^+\cup {\cal H}^-$
has two components, the  future,  ${\cal H}^+$, and the past ${\cal H}^-$  event horizons.
The future light cone of any point on ${\cal H}^+$ is tangent to ${\cal H}^+$ and is directed inside the black hole. Correspondingly, past light cones
on ${\cal H}_-$ are directed inside the white hole. A detailed discussion of this can be found in \cite{Misner:1973prb},\cite{Frolov:1998wf}. 

One can consider observers which rotate with respect to the Boyer–Lindquist coordinate grid (and therefore with respect to objects at the spatial infinity) with an angular coordinate velocity $d\varphi/ dt=-{g_{t\varphi} / g_{\varphi \varphi}} $. An important 
property of (\ref{2.1}) is that the only possible value for the angular velocity, when $r$ approaches $r_H$, is
\begin{equation}\label{2.7}
\Omega_H={a \over a^2+r^2_H}~~.
\end{equation}
The parameter $\Omega_H$ is called the angular velocity of the horizon.

Although Kerr solution (\ref{2.1})-(\ref{2.4}) looks complicated only
few features of the near horizon geometry are required for studying quantum effects we are interested in. These features are related to the structure 
of time-like isometries and properties of the so called Killing observers. A vector field $\zeta^\mu(x)$ on a manifold 
$\cal M$ is called a Killing field if it generates isometries of $\cal M$. The Killing field obeys the Killing equation 
\begin{equation}\label{2.8}
\zeta_{\mu;\nu}+\zeta_{\nu;\mu}=0~~.
\end{equation}
For the Kerr solution there is a distinguished Killing vector field, $\zeta=\partial_t-\Omega_H\partial_\varphi$, which is null on the horizon
\begin{equation}\label{2.9}
\zeta^2\mid_{\cal H}=0~~.
\end{equation}
Since $\cal H$ is the null hypersurface Eq. (\ref{2.9}) implies that $\zeta$ is a normal vector to $\cal H$. Properties of null hypersurfaces say that 
integral lines of $\zeta$ on $\cal H$ are geodesics:
\begin{equation}\label{2.10}
\nabla_\zeta \zeta=k \zeta~~.
\end{equation}
One can show that $\partial_\zeta k =0$ on $\cal H$, and, as a consequence, there is a 2D section, $\cal B$ of $\cal H$ where the Killing field is zero,
\begin{equation}\label{2.11}
\zeta^\mu\mid_{{\cal B}\in {\cal H}}=0~~.
\end{equation} 
The parameter $k$ in (\ref{2.10}) is called the surface gravity of the horizon, the section $\cal B$ is called the bifurcation surface of the Killing horizons.
Examples of Killing fields with bifurcating horizons are shown on Fig. (\ref{fig2}) for the case of Minkowsky space-time and eternal Schwarzschild black hole geometry. 

The Killing field $\zeta$ is time-like in the wedge to the right from ${\cal H}^+$ and ${\cal H}^-$. 
In this region one can define a frame of reference of observers whose 4-velocities $u^\mu$ are directed along $\zeta$, 
\begin{equation}\label{2.12}
u^\mu=\zeta^\mu/\sqrt{B}~~,~~B=-\zeta^2=-g_{tt}-2\Omega_Hg_{t\varphi}-\Omega^2_Hg_{\varphi\varphi}~~.
\end{equation}
Such observers are called  the Killing observers.
For a rotating black hole the given frame of reference is defined in a domain close to ${\cal H}$, where $B>0$. 
The congruence of the trajectories  is 
specified \cite{Hawking:1973uf} by 
the acceleration $w^\mu$, the rotation tensor $A_{\mu\nu}$ and the local angular velocity $\Omega(x)$
\begin{equation}\label{2.13}
w^\mu=\nabla_u u^\mu~~,~~
A_{\mu\nu}=\frac 12 h_\mu^\lambda h_\nu^\rho (\nabla_\rho u_{\lambda}-\nabla_\lambda u_{\rho})~~,
~~\Omega(x)=\left(\frac 12 A_{\mu\nu}A^{\mu\nu}\right)^{1/2}~~,
\end{equation}
where $h_\mu^\lambda=\delta_\mu^\lambda+u_\mu u^\lambda$.  Quantities (\ref{2.13}) appear under study of quantum systems in thermal equilibrium 
with the black hole, see Sec. \ref{HT}. Local angular velocity determines rotation of the Killing frame  with respect to a local inertial frame.

One can use (\ref{2.10}) to relate definition of the surface gravity
to the strength of gravity near the horizon,
\begin{equation}\label{2.14}
k=\lim_{r\to r_H}\left(\sqrt{Bw^2}\right)={r_H-MG \over a^2+r_H^2}~~.
\end{equation}
The right hand side (r.h.s.) of (\ref{2.14}) follows from  (\ref{2.2})-(\ref{2.4}), (\ref{2.12}).

It is convenient to change in the Boyer–Lindquist coordinates 
$\varphi$ to $\varphi'=\varphi - \Omega_H t$. In the new coordinates the Killing vector field is $\zeta=\partial_t$, that is, the Killing observers
do not move with respect to the new coordinate grid. Metric  (\ref{2.1}) can be rewritten as 
\begin{equation}\label{2.15}
ds^2=-B(dt+a_idx^i)^2+h_{ij}dx^idx^j~~,
\end{equation}
where $x^i=(r,\varphi',\theta)$, $a_idx^i=a_\varphi d\varphi'$. The non-vanishing components of acceleration and rotation are
$w_i=(\ln B)_{,i}/2$, $A_{ij}=\sqrt{B}(a_{i,j}-a_{j,i})/2$.

One can check that at small $r-r_H$
\begin{equation}\label{2.17}
B\simeq 4k^2h~r_H(r-r_H)~~,
~~h_{rr}\simeq {h r_H \over r-r_H}+O((r-r_H)^2)~~,~~h(\theta)\equiv{\Sigma(r_H,\theta) \over 2r_H\sqrt{(MG)^2-a^2}}~.
\end{equation}
Here we took into account that $h_{rr}=g_{rr}$. It is convenient to introduce a new coordinate $\rho$:
\begin{equation}\label{2.18}
r-r_H\simeq {\rho^2 \over 4r_H}~~.
\end{equation}
connected with the proper distance $L$ to the horizon 
\begin{equation}\label{2.16}
L(r,\theta)=\int_{r_H}^rdr'\sqrt{h_{rr}} \simeq \sqrt{h} \rho~~.
\end{equation}

In the leading approximation
$B\simeq hk^2\rho^2$. Since the local angular velocity 
vanishes near the horizon, $\Omega=O(\rho)$, terms $a_idx^i$ in (\ref{2.15}) can be neglected near $\cal H$. 

One comes to the following form of near-horizon black hole metric (\ref{2.1}) :
\begin{equation}\label{2.19}
ds^2\simeq h(\theta) (- k^2\rho^2 dt^2+d\rho^2)+dl^2~~,
\end{equation}
\begin{equation}\label{2.20}
dl^2=\gamma_{\theta\theta}d\theta^2+\gamma_{\varphi\varphi}(d\varphi')^2~~,
\end{equation}
where $\gamma_{\alpha\beta}=h_{\alpha\beta}(r=r_H)$ and (\ref{2.20}) is the metric on ${\cal B}$.  It can be shown that
$\gamma_{\theta\theta}=g_{\theta\theta}(r_H)$, $\gamma_{\varphi\varphi}=g_{\varphi\varphi}(r_H)$.
Therefore the space-time near $\cal H$ has the product structure $R^2\times {\cal B}$. In case of the Schwarzchild black hole ${\cal B}$ is $S^2$, $\cal B$ is closed and has the topology of $S^2$ for the Kerr solution, and ${\cal B}=R^2$ for flat space-time. 

The area $\cal A$ of ${\cal B}$,
\begin{equation}\label{2.21}
{\cal A}=\int d\theta d\varphi \sqrt{\det \gamma_{\alpha\beta}}=4\pi(r_H^2+a^2)~~,
\end{equation}
is called the area of black hole horizon. It plays an important role in the subsequent discussions.
To get (\ref{2.21}) one should use (\ref{2.2}),(\ref{2.3}).

\section{Classical fields, quantization, quasiparticles}\label{fields}

Since stationary black holes have a universal structure near the horizon it is worth studying  classical 
fields in this region by using, as an example, a free scalar field $\phi$. The field equation is 
\begin{equation}\label{4.1}
(\nabla^2-m^2)\phi={1 \over \sqrt{-g}}\partial_\mu(\sqrt{-g} g^{\mu\nu}\partial_\nu \phi) -m^2\phi=0~~~,
\end{equation}
where $m$ is the mass of the field.  Analysis of wave equations like (\ref{4.1}) on the Kerr background can be found in
\cite{Chandrasekhar:1985kt}.  According to (\ref{2.19}), when $r$ close $r_H$,  Eq. (\ref{4.1}) reduces to:
\begin{equation}\label{4.2}
\left[-{1 \over k^2 \rho^2}\partial_t^2+\partial_\rho^2\right]\phi-L\phi=0~~.
\end{equation}
Here $L=h(\Delta+m^2)$ and $\Delta$ is the Laplace operator on $\cal B$.  Solutions to (\ref{4.2}),
which are interpreted by a Killing observer as excitations with energies $\omega$, are
\begin{equation}\label{4.3}
i\partial_t\phi=\omega \phi~~.
\end{equation}
If $\psi_\lambda (\theta, \varphi)$ is an eigen-function of $L$ with an eigen-value $\lambda$, solution to (\ref{4.2}), (\ref{4.3}) 
can be written as
\begin{equation}\label{4.4}
\phi(t,\rho,\theta,\varphi)=e^{-i\omega t}\phi_\omega(x)\psi_\lambda (\theta, \varphi)~~,
\end{equation}
\begin{equation}\label{4.5}
(-\partial_x^2+V_\lambda(x))\phi_\omega(x)=\omega^2\phi_\omega(x)~~.
\end{equation}
Here $\rho=Ce^{kx}$ and $V_\lambda(x)=\lambda C^2k^2e^{2kx}$, $C$ is a constant. 
Note that $\cal B$ is compact and $h$ is finite on $\cal B$. One can show that 
eigenvalues $\lambda$ of $L$ are discrete and positive.

The spectrum of $\omega$, called the single-particle energies, is defined  by a Schroedinger-like equation  (\ref{4.5}) with an effective potential
$V_\lambda$. We assume that $C>0$ and imply that limit $\rho\to 0$ corresponds to $x\to -\infty$. A straightforward conclusion from  (\ref{4.5})  is that
fields near the horizon  are effectively massless and the spectrum of $\omega$ is continuous. Effects related to non-zero mass $m$ or properties of $L$ are exponentially suppressed.

As an illustration, we give an exact form of $V_\lambda$ for the Schwarzschild black hole ($a=0$)
\begin{equation}\label{4.6}
V_\lambda(x)=B(r)\left(m^2+{\partial_r B(r)\over r}+{l(l+1) \over r^2}\right)~~,~~x=\int^r {dr' \over B(r')}~~,
\end{equation}
$B(r)=1-r_H/r$.
Eigein-values $\lambda =l(l+1)$ belong to the spectrum of a Laplacian on unit 2-sphere. At $r\to r_H$ (\ref{4.6}) reduces to (\ref{4.5}).
Near the horizon, $x\to -\infty$, the potential is exponentially small. The mass of the field dominates, $V_\lambda(x)\simeq m^2$,
as  $x\to +\infty$. Potential $V_\lambda(x)$ reaches a maximum at some point $x$ whose position depends on $l$.

Since we are interested in near-horizon physics the above analysis suggests a serious technical simplification: one can focus only on dynamics
in the $(t,r)$ coordinate sector where all fields are effectively massless.

For further purposes we recall basic elements of quantum field theory on classical backgrounds, for details see \cite{Fursaev:2011zz}. Take, 
as an example,
the free scalar field with equation (\ref{4.1}).  For a pair  of solutions, $f_1$, $f_2$, to  (\ref{4.1}) one introduces a relativistic inner product
\begin{equation}\label{4.7}
\langle f_1,f_2\rangle=\int_{\Sigma}d\Sigma^\mu~i(f_1^\star \partial_\mu f_2- \partial_\mu  f_1^\star f_2)~~,
\end{equation}
which is taken on a Cauchy hypersurface $\Sigma$. The choice of $\Sigma$ is unimportant since (\ref{4.7}) conserves under variations of $\Sigma$.
Suppose that $f_i^{(\pm)}$ are solutions enumerated by a certain set of indices $i$, discrete or continuous, and normalized as
\begin{equation}\label{4.8}
\langle f_i^{(\pm)},f_j^{(\pm)}\rangle=\pm \delta_{ij}~~,~~\langle f_i^{(+)},f_j^{(-)}\rangle=0~~.
\end{equation}
Suppose also that any solution to  (\ref{4.1}) can be written as a linear combination  
\begin{equation}\label{4.9}
\phi(x)=\sum_ia_i  f_i^{(+)}(x)+\sum_j b_j^\star  f_j^{(-)}(x)~~.
\end{equation}
Quantization procedure implies that $\phi$, $a_i$ and $b_j^\star$ are replaced 
with operators $\hat{\phi}$, $\hat{a}_i$ and $\hat{b}_j^+$ with the 
following commutation relations:
\begin{equation}\label{4.10}
[\hat{a}_i,\hat{a}_j^+]=\delta_{ij}~~,~~[\hat{b}_i,\hat{b}_j^+]=\delta_{ij}~~.
\end{equation}
Operators $\hat{a}_j^+,\hat{b}_j^+$ create quasiparticles which, in general, may not carry any definite energy. In stationary space-times
with the Killing field $\zeta=\partial_t$ there is a special set of single-particle modes such as (compare with (\ref{4.3}))
\begin{equation}\label{4.11}
i\partial_t f_i^{(\pm)}=\pm \omega_i^{(\pm)}f_i^{(\pm)}~~.
\end{equation}
$\omega_i$ are positive numbers we call single-particle energies. Correspondingly, the Killing observers interpret $f_i^{(\pm)}$ as 
excitations with certain energies. Spectrum of single-particle energies can be found from equations like Eq. (\ref{4.5}).

In General Relativity definition of a particle depends on the frame of reference where observations are done. To see how 
different choices of creation and annihilation operators are connected suppose that $\phi$ is a real field. One has two decompositions
\begin{equation}\label{5.1}
\hat{\phi}(x)=\sum_i(\hat{a}_i  f_i^{(+)}(x)+ \hat{a}_i^+  f_i^{(-)}(x))=\sum_j(\hat{\bar{a}}_j  \bar{f}_j^{(+)}(x)+ \hat{\bar{a}}_j  ^+  
\bar{f}_j^{(-)}(x))
\end{equation}
corresponding to different definitions of quasiparticles,  $\hat{a}_i$, $\hat{\bar{a}}_j$. Reality condition implies that 
$f_i^{(-)}=(f_i^{(+)})^\star$, $\bar{f}_i^{(-)}=(\bar{f}_i^{(+)})^\star$. By using normalization conditions (\ref{4.8}) one finds from (\ref{5.1}) 
that 
\begin{equation}\label{5.2}
\hat{a}_i=\langle f_i^{(+)} , \hat{\phi}\rangle=\sum_j(c_{ij}\hat{\bar{a}}_j +d_{ij}\hat{\bar{a}}_j ^+)~~,
\end{equation}
\begin{equation}\label{5.3}
c_{ij}=\langle f_i^{(+)} ,  \bar{f}_j^{(+)}\rangle~~,~~d_{ij}=\langle f_i^{(+)} ,  \bar{f}_j^{(-)}\rangle~~.
\end{equation}
Coefficients $c_{ij}$, $d_{ij}$ are called the Bogoliubov coefficients.  As follows from (\ref{5.2}) the average number of quasiparticles created by
$\hat{a}_i$ is non-trivial, in general,
\begin{equation}\label{5.4}
N_i=\langle \bar{0}|\hat{a}_i^+\hat{a}_i|\bar{0}\rangle=\sum_j |d_{ij}|^2~~,
\end{equation}
in the vacuum state $|\bar{0}\rangle$ which does not contain particles of the other sort,  $\hat{\bar{a}}_j|\bar{0}\rangle=0$.

\section{Quantum evaporation of a black hole}\label{evap}

We present now a sketch of arguments demonstrating the Hawking effect of quantum evaporation of a black hole \cite{Hawking:1975vcx}.  
The effect is related to physics near the black hole horizon where, according to Sec. \ref{def}, the geometry has the universal form, see Eq. (\ref{2.19}).
In the leading approximation the dynamics occurs in $(t,r)$ or $(t,\rho)$ coordinate sector, while dependence on angles $\theta$ and $\varphi$
is unimportant.

For simplicity we consider the behavior of quantum scalar field $\phi$
outside a spherically symmetric star which collapses and creates a Schwarzschild black hole. We assume that $\phi$ does not depend on 
the angles, that is, $\phi$ is the so called $s$-mode. Since mass of the field is not
important near the horizon we also assume that $\phi$ is massless. 

Wave equation (\ref{4.1}) reduces to $\nabla^2\phi=0$, where $\nabla^2$ is the operator on a 2D spacetime ${\cal M}_2$.
Outside the star 
\begin{equation}\label{5.5}
ds^2= - B(r)dt^2+{dr^2 \over B(r)}~~,~~B(r)=1-{r_H \over r}~~.
\end{equation}
A standard approach is to introduce ingoing, $(v,r)$, and outgoing, $(u,r)$, 
Eddington-Finkelstein coordinates
\begin{equation}\label{5.8a}
u=t-x~~,~~v=t+x~~,~~x=\int^r{dr' \over B(r')}~~,
\end{equation}
\begin{equation}\label{5.8b}
ds^2=-Bdv^2+2dvdr=-Bdu^2-2dudr=-B(r)dudv~~.
\end{equation}
Null coordinates $u$, $v$ are retarded and advanced times, respectively. Lines of constat $u$ or $v$ are radial rays.
Past-directed null geodesics end up on the past null infinity denoted by $\mathcal{I}^{-}$, 
future-directed null geodesics  end up on the future null infinity $\mathcal{I}^{+}$. These infinities can be used as parts of the Cauchy surface $\Sigma$  where relativistic product (\ref{4.7}) is defined. In computations $\mathcal{I}^{+}$ and $\mathcal{I}^{-}$
are replaced, respectively, with null surfaces $v=C>0$ or $u=-C<0$, with large $C$.
For an eternal black hole shown on Fig. \ref{fig2} coordinates  $(v,r)$
are continued across ${\cal H}^+$ inside the black hole, the outgoing coordinates can be continued across ${\cal H}^-$ .

2D massless scalar fields are easy to analyze since ${\cal M}_2$ is conformally flat.
The metric can be brought to the form $ds^2=-e^{2\sigma}dudv$, and a general solution to the wave equation 
\begin{equation}\label{5.6}
\nabla^2\phi= -4e^{-2\sigma}\partial_u\partial_v \phi=0~~
\end{equation}
is a combination of left-moving and right-moving modes, $\phi(u,v)=\phi_R(u)+\phi_L(v)$. 

\begin{figure}
		\includegraphics[width=10cm]{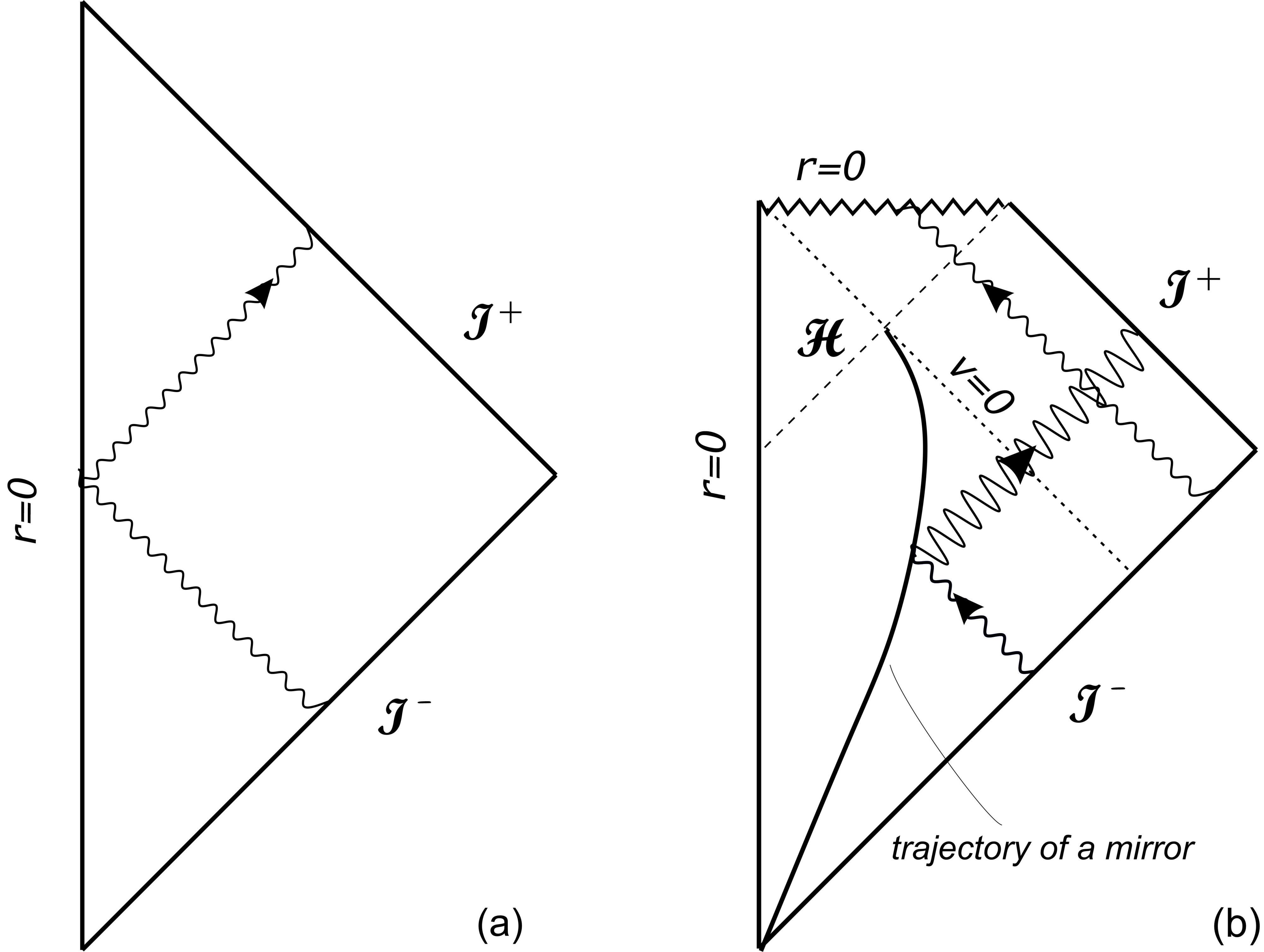} 
		\caption{Carter-Penrose diagrams which show propagation of waves from $\mathcal{I}^{-}$  in Minkowsky space-time, ($a$), and in space-time of a collapsing star, ($b$). In flat space-time all modes from $\mathcal{I}^{-}$ reach 
$\mathcal{I}^{+}$ unchanged.  The right figure shows the "escape" modes 
$\phi_{\mbox{\small{esc}}}(u,v)$ defined at $v<0$. These modes leave the star just before the horizon is formed and behave as if reflected from an imaginary mirror. Modes $\phi_H(v)$
are shown as waves from $\mathcal{I}^{-}$ at $v>0$. They cross $\cal H$ and end on the singularity $r=0$.}
		\label{fig1}
\end{figure}

Before we proceed with solutions on black hole geometries it is instructive to consider fields on 2D Minkowsky space-time 
with metric (\ref{5.8b}), where $B=1$, $x=r$ and $r>0$. A complete set of modes is
\begin{equation}\label{5.10a}
f^{(+)}_{\omega}(u,v)={1 \over \sqrt{4\pi \omega}}\left(e^{-i\omega v}-e^{-i\omega u}\right)~~,~~\langle f^{(+)}_{ \omega},f^{(+)}_{\sigma}\rangle=\delta(\omega-\sigma)~~.
\end{equation}
Since $r$ is a radial coordinate with the center at $r=0$ each mode is a combination of left-moving and right-moving waves with the Dirichelt boundary condition,
$f^{(+)}_{\omega}=0$, at $r=0$. In these coordinates waves coming
from $\mathcal{I}^{-}$ are reflected from the center and travel, unchanged, to $\mathcal{I}^{+}$, see Fig. \ref{fig1}. Due to this reflection condition
null infinities $\mathcal{I}^{\pm}$
are the Cauchy surfaces where the relativistic inner product $\langle f^{(+)}_{ \omega},f^{(+)}_{\sigma}\rangle$
can be defined as
\begin{equation}\label{5.7}
\langle f_1,f_2\rangle=i\int_{\mathcal{I}^{-}}dv~(f_1^\star \partial_v f_2- \partial_v  f_1^\star f_2)=
i\int_{\mathcal{I}^{+}}du~(f_1^\star \partial_u f_2- \partial_u  f_1^\star f_2)~~.
\end{equation}
Formula (\ref{5.7}) also holds on space-times with metric $ds^2=-e^{2\sigma}dudv$, it does not depend $\sigma$.

There is a principle distinction between fields near a collapsing star and fields in the flat space-time:  for the star there are 
ingoing waves, $\phi_H(v)$, which cannot escape to $\mathcal{I}^{+}$. Such waves come after the black hole horizon is formed, 
cross $\cal H$ and move inside the black hole. Schematically the 2D part of a collapsing star is shown on Fig. \ref{fig1}.
A general solution to (\ref{5.6}) in the null coordinates 
is a combination of left and right-moving modes of the following form:
\begin{equation}\label{5.7a}
\phi(u,v)=\phi_{\mbox{\small{esc}}}(u,v)+\phi_H(v)~~.
\end{equation}
By $\phi_{\mbox{\small{esc}}}(u,v)$ we denote waves which can escape to $\mathcal{I}^{+}$. Like waves in the Minkowsky space-time
$\phi_{\mbox{\small{esc}}}(u,v)$ satisfy a sort of reflection condition. 
If $v=0$ is the trajectory of the last ingoing ray which escapes the black hole, just before
the horizon is formed, $\phi_{\mbox{\small{esc}}}(u,v)$, $\phi_H(v)$ are defined for $v<0$ or $v>0$, respectively.
For solutions (\ref{5.7a}) the equivalent Cauchy surfaces are $\mathcal{I}^{-}$, in far past, or $\mathcal{I}^{+}\cup {\cal H}$, in future.

To determine $\phi_{\mbox{\small{esc}}}$ one can trace modes inside the collapsing star and find the ``boundary condition''. This option is
complicated and requires additional model assumptions.  
Another strategy is to define $\phi_{\mbox{\small{esc}}}$ by requiring its regularity 
across $\cal H$.  The fact that $\phi_{\mbox{\small{esc}}}$ cannot be taken as waves  (\ref{5.10a}) in flat spacetime is clear since outgoung coordinate $u$ is not analytical on $\cal H$.  

There are alternative coordinates, the Kruskal coordinates,
\begin{equation}\label{5.9}
U=U(u)=-ae^{-ku}~~,~~V=V(v)=be^{kv}~~,
\end{equation}
which are both analytic on $\cal H$. Here $a,b$ are some positive constants which can be chosen such that  
$ds^2\simeq - dUdV$ near $\cal H$. The event horizon $\cal H$ is defined by conditions $U=0$, $v>0$. It is important that
$U$ and $V$ can be interpreted near $\cal H$ as retarded and advanced times for freely falling observers .

In the black hole exterior $U>0$, and $U$ can be continued inside the black hole, where $U<0$. Therefore, for example, the following modes:
\begin{equation}\label{5.10}
f^{(+)}_{\mbox{\small{in}}, \omega}(u,v)={1 \over \sqrt{4\pi \omega}}\left(e^{-i\omega v}-e^{-i\omega  U(u)}\right)~~,~~-\infty <u,v< \infty~~
\end{equation}
behave well at $\cal H$. These modes: 

i) are ingoing waves $e^{-i\omega v}$ with frequency $\omega$ as measured by freely moving observers in the far past, near 
$\mathcal{I}^{-}$;

ii) are outgoing waves $e^{-i\omega U}$ with frequency $\omega$  for freely falling observers near $\cal H$;

iii) make a complete set on the Cauchy surfaces $\mathcal{I}^{-}$ or $\mathcal{I}^{+}\cup {\cal H}$. 

Therefore for the collapsing star a quantum state which is experienced as a vacuum by freely falling observers,
including observers near  $\mathcal{I}^{-}$, should be modeled by a state $|\Psi \rangle$
determined by the condition: 
\begin{equation}\label{5.13}
\hat{a}_\omega|\Psi\rangle=0~~,
\end{equation}
where operators $\hat{a}_\omega$ are introduced
by (\ref{4.9}) with respect to modes $f^{(+)}_{\mbox{\small{in}}, \omega}$ defined in (\ref{5.10}).

The $\phi_H$ part of modes (\ref{5.10}) is simply proportional to $e^{-i\omega v}$ for $v>0$. The escape 
part, $\phi_{\mbox{\small{esc}}}$,  $v<0$, looks as a solution to the wave equation 
in the presence of a perfectly reflecting accelerated mirror which moves along the trajectory
\begin{equation}\label{5.11}
v=U(u)~~,
\end{equation}
see Fig. \ref{fig1}.
This fictitious mirror plays the role of a strong gravity, it transforms ingoing waves $e^{-i\omega v}$ in the escape part  to 
outgoing waves $e^{-i\omega  U(u)}$.

It is well-known that accelerated mirrors create particles. Analogously the collapsing body creates the flux of the Hawking radiation. To see this define modes 
\begin{equation}\label{5.12}
f^{(+)}_{\mbox{\small{out}}, \sigma}(u,v)={1 \over \sqrt{4\pi \sigma}}\left(e^{-i\sigma u}-e^{-i\sigma \bar{v}(v)}\right)~~,~~-\infty <u< \infty~~,
~~v< 0~~,
\end{equation}
where $\bar{v}(v)=-{1 \over k}\ln(-\frac{v}{a})$. These out-modes:

i) are reduced to outgoing waves $e^{-i\omega u}$ with frequency $\omega$ as measured by freely moving observer in the far future, 
near 
$\mathcal{I}^{+}$;
 
ii) satisfy the Dirichlet condition on "trajectory" (\ref{5.11});

iii) do not have $\phi_H(v)$ part;

iv) make a complete set with the same normalization on the null surfaces discussed above. 

Define creation and annihilation operators 
$\hat{b}^+$ and $\hat{b}$ for out-modes $f^{(+)}_{\mbox{\small{out}}, \sigma}$.  The corresponding quasiparticles are interpreted by observers
near $\mathcal{I}^{+}$ as particles with certain energies.  One identifies these particles with the Hawking quanta.

The number of the Hawking quanta in given state is, see Eqs. (\ref{5.3}),(\ref{5.4}),
\begin{equation}\label{5.14}
N_\sigma =\langle \Psi| \hat{b}^+_\sigma\hat{b}_\sigma| \Psi\rangle=\int d\omega |d_{\sigma,\omega}|^2~~,
\end{equation}
\begin{equation}\label{5.15}
d_{\sigma,\omega}=\langle f^{(+)}_{\mbox{\small{out}}, \sigma},f^{(-)}_{\mbox{\small{in}}, \omega}\rangle=-{1 \over 2\pi}
\sqrt{\sigma \over \omega}\int^\infty_{-\infty} du ~e^{i\sigma u+i\omega U(u)}~~.
\end{equation}
The product in (\ref{5.15}) is defined on $\mathcal{I}^{+}$. Integral in r.h.s. of (\ref{5.15}) can be performed
after the substitution $t=e^{-ku}$
\begin{equation}\label{5.16}
d_{\sigma,\omega}=-{1 \over 2\pi k}\sqrt{\sigma \over \omega}\exp\left(-{\pi \sigma \over 2k}\right)\Gamma\left(-{i\sigma \over k}\right)
(\omega a)^{i\sigma \over k}~~,
\end{equation}
where $\Gamma(z)$ is the $\Gamma$-function
\begin{equation}\label{5.17}
\Gamma(z)=\int_0^\infty dt~t^{z-1}e^{-t}~~.
\end{equation}
By using the property $|\Gamma(iz)|^2=\pi/(z\sinh \pi z)$ one gets the number of the Hawking quanta in the form:
\begin{equation}\label{5.18}
N_\sigma={\Gamma_\sigma \over e^{\sigma /T_H}-1}~~.
\end{equation}
Coefficients $\Gamma_\sigma$ are called the grey-body factors. In the considered case $\Gamma_\sigma$
is a constant which includes a regularized integral over $\omega$. 

A remarkable fact is that the Hawking quanta are distributed  according to Planck's law with a temperature 
\begin{equation}\label{3.2}
T_H={k \over 2\pi}~~
\end{equation}
called the Hawing temperature.

The simplified analysis presented here can be extended beyond the $s$-mode approximation and near-horizon approximation. Massless modes which depend on angles will experience a partial reflection on the potential $V_l(r)$, see (\ref{4.6}). This effect yields non-trivial
factors $\Gamma_\sigma$.

The Hawking effect can be interpreted as a process of creation of particle-antiparticle pairs in the strong gravitational field
near the horizon. Antiparticles created in this process tunnel inside the black hole, while particles make the Hawking flux. Semiclassical estimations of the tunneling
probability  \cite{Parikh:1999mf} are in agreement with (\ref{5.18}), see \cite{Vanzo:2011wq} for a review.

If, after the collapse of the star, the black hole evaporates completely, it results in violation of the unitarity and information loss since the initial pure state 
$|\psi\rangle$ evolves to a mixed thermal state. This paradox, despite several interesting hypothesis \cite{Susskind:1993if}, has not been 
resolved so far.

\section{Thermodynamic laws of black holes}\label{noether}
\subsection{Black hole mechanics}

If a black hole appears as a result of the gravitational collapse
of a star, it quickly reaches a stationary state characterized
by a
certain mass $M$ and an angular momentum $J$.  By using purely classical Einstein equations, or on the base of definitions of
(\ref{2.7}), (\ref{2.14}), (\ref{2.21}),
one arrives at the following variational formula \cite{Bardeen:1973gs}:
\begin{equation}\label{5.20}
\delta M=T_H \delta S^{BH}
 +\Omega_H\delta J~~,
\end{equation}
where $T_H$ is given by  (\ref{3.2}) and
\begin{equation}\label{5.21}
S^{BH}={1 \over 4G}{\cal A}~~.
\end{equation}
${\cal A}$ is the surface area of the horizon (see above) and
$G$ is the Newton gravitational constant.

The quantity $S^{BH}$
was introduced in \cite{Bekenstein:1972tm}-\cite{Bekenstein:1974ax},\cite{Hawking:1975vcx}
and is called the Bekenstein-Hawking entropy.

Relation (\ref{5.20}) has the form of the
first law of thermodynamic where
$S^{BH}$ has the meaning of an entropy,
$T_H$ is a temperature, and $M$ is an internal energy.  
If the collapsing matter
was not electrically neutral a black hole has an additional
parameter, an electric charge $Q$. Then the r.h.s. of (\ref{5.20}) would acquire 
additional term $\Phi_H \delta Q$, where
$\Phi_H$ is the difference of the electric potential at the horizon
and at infinity. $M,J,Q$ are the only
parameters a black hole in the Einstein-Maxwell
theory can have.
Its metric in the most general case is the Kerr-Newmann
metric. This statement is known as the "no-hair"
theorem, see e.g. \cite{Frolov:1998wf}.

The Bekenstein-Hawking entropy is one of the most misterious quantities in black hole thermodynamics. 
For super-massive black holes with masses of the order
of $10^9$ solar masses $S^{BH}$ is of the order of $10^{95}$, 
it is eight orders of magnitude larger than the entropy
of the microwave background radiation in the visible part of the Universe.
This raises a natural question about microscopic
degrees of freedom whose number is consistent with the
Bekenstein-Hawking entropy.

The reason why this question is fundamental is because it
goes beyond the black hole physics itself. On one hand, its answer  may
give important insights into the as yet mysterious nature of
quantum gravity. On the other hand, since the thermodynamics of black holes is a low-energy phenomenon,
understanding of the black hole entropy may be possible without
knowing details of quantum gravity, for example in the framework of the perturbative quantum gravity methods.

\subsection{Black holes and Euclidean theory}\label{eucl}

There is another way to see that black holes look as thermodynamic systems.  Consider the partition function 
of a quantum system with a (normally ordered) Hamiltonian $:\hat{H}:$ at temperature $T=\beta^{-1}$,
\begin{equation}\label{5.30}
Z(\beta)=\mbox{Tr}~e^{-\beta :\hat{H}:}~~~.
\end{equation}
As is known, (\ref{5.30}) can be interpreted as a trace of the evolution operator $\hat{U}(t)=e^{-it\hat{H}}$  with imaginary time interval $t=-i\beta$. This allows one (see discussion in next sections) to represent $Z(\beta)$ as a path integral in the corresponding Euclidean quantum theory with periodic 
or antiperiodic boundary conditions in the Euclidean time $\tau=it$. In application to black holes in vacuum this implies that instead of the Lorentzian Ricci flat
solutions, $R_{\mu\nu}=0$, see (\ref{i.1}), we should consider analogous solutions on Riemannian, or Euclidean manifolds with 
the signature $+,+,+,+$. Such solutions are called gravitational instantons. 

Consider the Kerr solution (\ref{2.1})-(\ref{2.4}).  The corresponding instanton can be obtained from (\ref{2.1})-(\ref{2.4}) by the Wick rotation of
time, $t=-i\tau$ and by changing $a$ to $-ia$ to ensure that non-diagonal term $2g_{t\varphi}dt d \varphi$ in metric (\ref{2.1}) remains real.
(It should be noted that quantum theory which we discuss in next sections does not require the Wick rotation of $a$, so one can work in principle
with complex metrics.)
By using parameters of the Lorentzian solution, $r_H$, $\Omega_H$, $k$ one defines analogous parameters $\bar{r}_H=r_H(ia)$, 
$\bar{\Omega}_H=\Omega_H(ia)$,  $\bar{k}=k(ia)$ for the instanton. The Euclidean Kerr solution has analogous symmetries. 
The Killing vector field $\zeta=\partial_\tau+\bar{\Omega}_H\partial_\varphi$ does not have a horizon, as in the Lorentzian theory,
but it has fixed points located on a closed 2D surface $r=\bar{r}_H$, the Euclidean horizon, 
which we denote by $\cal B$, in the same way as the bifurcation surface. Near $\cal B$ the Euclidean metric looks as follows
(compare with (\ref{2.19})):
\begin{equation}\label{5.31}
ds^2\simeq \bar{h}(\theta) (\bar{k}^2\rho^2 d\tau^2+d\rho^2)+dl^2~~,~~0<\tau\leq \beta~~.
\end{equation}
At arbitrary periodicity (\ref{5.31}) has a conical singularity at $\rho=0$ ($r=\bar{r}_H$). The singularity disappears if 
\begin{equation}\label{5.32}
\beta=\beta_H={2\pi \over \bar{k}}~~.
\end{equation} 
Thus, the regularity condition requires that the period $\beta_H$ coincides with the inverse Hawking temperature (\ref{3.2}).

The fact that the Euclidean horizon is a fixed point set of the Killing field associated to time translations also implies a ``thermodynamic'' form of the gravity action on the black hole instanton. It is easy to check that the Euclidean action, $I_E[\phi]$, say, for a scalar field $\phi$, between a constant time 
hypersurface $\Sigma_\tau$  and a hypersurface $\Sigma_{\tau'}$  with $\tau'=\tau+\beta$ has the form 
$I_E[\phi]=\beta H$, where $H$ has a form of the Hamiltonian of the system. On static solutions, $\partial_\tau\phi=0$, $H$ coincides with the energy of the system. 

Consider now the Einstein-Hilbert action on Riemannian (Euclidean) manifolds $\cal M$ 
\begin{equation}\label{5.33}
I_E[g]=- {1 \over 16\pi G}\left[\int_{\cal M} d^4x \sqrt{g}~ R+2\int_{\partial {\cal M}} d^3y \sqrt{h}~K\right]~~.
\end{equation} 
The last term in the r.h.s. of (\ref{5.33}) should be added, according to Gibbons and Hawking \cite{Gibbons:1976ue}, when $\cal M$ has a boundary $\partial {\cal M}$. This term depends on the trace $K$ of the extrinsic curvature of  $\partial {\cal M}$ and it guarantees that variations $I_E[g]$ do not contain variations of normal derivatives of the metric on  $\partial {\cal M}$. To avoid infrared 
divergences in (\ref{5.33}) on asymptotically flat space-times, $I_E[g]$ is defined with a subtraction of the corresponding action
on a flat space-time.

If the Killing field $\partial_\tau$ does not have fixed points
the only boundaries of constant $\tau$ hypersurfaces belong to $\partial {\cal M}$. Then action (\ref{5.33}) computed
between $\Sigma_\tau$  and  $\Sigma_{\tau+\beta}$ has the structure, $I_E=\beta H$, where $H$ is a Hamiltonian. On asymptotically flat 
solutions, after the subtraction in $I_E[g]$, $H$ coincides with the ADM mass \cite{Hawking:1995fd}.

Situation is different for black hole instantons since $\partial_\tau$ has fixed points on $\cal B$. As a result, $\Sigma_\tau$
end on $\cal B$, and the Euclidean horizon becomes an internal boundary. To calculate the action on $\cal M$
one should consider the near-horizon part  ${\cal M}_\epsilon$ of $\cal M$ separately. The metric of ${\cal M}_\epsilon$ 
can be approximated by (\ref{5.31}) with the outer boundary located at $\rho=\epsilon$. The 
gravitational action taken on a black hole instanton then becomes  \cite{Gibbons:1976ue}\cite{Hawking:1995fd} 
\begin{equation}\label{5.34}
I_E=\beta_H (M-\Omega_HJ)-{{\cal A} \over 4G}~~,
\end{equation}
where $\cal A$ is the area of $\cal B$.
The first term in the r.h.s. of (\ref{5.34}) appears from a domain outside of ${\cal M}_\epsilon$ where constant time hypersurfaces
do not have intersections. The last term, $-{\cal A} / (4G)$ appears from the boundary term,  
\begin{equation}\label{5.35}
\lim_{\epsilon \to 0}{1 \over 8\pi G}\int_{\partial {\cal M}_\epsilon} d^3y \sqrt{h}~K={{\cal A} \over 4G}~~,
\end{equation} 
see discussion of this point, based on topological arguments, in \cite{Banados:1993qp} .

It follows from (\ref{5.34}) that $I_E/\beta_H$ can be interpreted as the free energy of a thermodynamic system 
with energy $M$, entropy ${\cal A} / (4G)$, and temperature $1/\beta_H$, in accord with first law (\ref{5.20}).

It will be important for the future discussion to point out that thermodynamic form of the action  (\ref{5.34}) can be extended to the case
when the Euclidean time $\tau$ in the instanton solution has an arbitrary period $\beta$. Such a geometry is not regular because of conical 
singularities at $\cal B$ with the deficit angle $2\pi(1-\beta/\beta_H)$. For this reason it is not a solution of the Einstein equations 
near $\cal B$. The black hole thermodynamics at $\beta\neq \beta_H$ is called an off-shell approach.
Components of the Riemann tensor on manifolds with conical singularities behave as distributions at $\cal B$, see \cite{Fursaev:1995ef}. In particular, the integral curvature is
\begin{equation}\label{5.36}
\int_{\cal M} d^4x \sqrt{g}~ R=4\pi \left(1-{\beta \over \beta_H}\right){\cal A}+\int_{{\cal M}/{\cal B}} d^4x \sqrt{g}~ R~~,
\end{equation} 
where ${\cal M}/{\cal B}$ is the regular part of $\cal M$. By using (\ref{5.35}), (\ref{5.36}) it is not difficult to check that the
off-shell action,
\begin{equation}\label{5.37}
I_E=\beta~(M-\Omega_HJ)-{{\cal A} \over 4G}=I_E(\beta,M,J)~~,
\end{equation}
holds the thermodynamic form.  The advantage of the off-shell formulation is that $\beta$ is a free parameter which is not related to
the mass $M$ and angular momentum $J=Ma$ of a black hole. If $I_E/\beta=F$ is interpreted as a free energy, the Bekenstein-Hawking entropy can be derived by using statistical-mechanical formula:
\begin{equation}\label{5.38}
S^{BH}=\beta^2\partial_\beta F(\beta,M,J)_{\beta=\beta_H}=\left(\beta \partial_\beta-1\right)I_{E,\beta=\beta_H}~~.
\end{equation}
We use definition (\ref{5.38}) in what follows.

To avoid unnecessary complications boundary conditions for a black hole have not been specified in the above analysis. 
The importance of boundary conditions has been pointed out in \cite{York:1986it} for the case of a Schwarzschild black hole
inside a spherical cavity of a finite radius. It can be shown by using the Gibbons-Hawking action that such a black hole at certain size of the cavity behaves as a stable  thermodynamic system.

Asymptotically anti-de Sitter black hole solutions in the Einstein gravity with a negative cosmological constant have a similar property \cite{Hawking:1982dh}: they can be in stable equilibrium state when their size is greater
than the radius of the anti-de Sitter space.

Thermodynamics of black holes in higher dimensional gravity theories with 
the negative cosmological constant is used  \cite{Witten:1998zw} to study
finite-temperature gauge theories in the framework of the so called AdS/CFT correspondence \cite{Witten:1998qj}, \cite{Maldacena:1997re} .
Interestingly, charged black holes in anti-de Sitter space-times have a phase structure similar to that
of the van der Waals-Maxwell liquid-gas systems in a space-time of one-dimension lower \cite{Chamblin:1999hg}.

\subsection{Black holes in generalized gravity theories}

It should be noted that variational formulas analogous to (\ref{5.20}) can be found for asymptotically flat black hole 
solutions in a general classical theory of gravity arising from a diffeomorphism invariant Lagrangian \cite{Wald:1993nt}. In these theories one comes to a 
generalization of the Bekenstein-Hawking formula (\ref{5.21}).  Moreover, the  black hole entropy can be interpreted as a Noether charge associated to the horizon Killing field $\zeta$. As an example, consider a modification of the Einstein gravity by terms quadratic in curvatures. The bulk part of gravity action (on Euclidean manifolds) is
\begin{equation}\label{5.22}
I[g,~(G,\Lambda, c_i)]=-\int d^4x \sqrt{g}
\left[ \frac{R}{16\pi G}-2\Lambda
-c_1 R^2 -c_2 R_{\mu\nu}R^{\mu\nu} -c_3
R_{\alpha\beta\mu\nu}R^{\alpha\beta\mu\nu}
\right]~~~,
\end{equation}
where $R_{\alpha\beta\mu\nu}$, $R_{\mu\nu}$, $R$ are the Riemann and Ricci tensor, as well as the scalar curvature, respectively.
Introduce 2 normal vectors $n_i^\mu$ at the bifurcation surface $\cal B$, $(n_i\cdot n_j)=\delta_{ij}$, and define at  $\cal B$ the following invariants:
\begin{equation}\label{5.23}
R_{ii}=R_{\mu\nu}n_i^\mu n_j^\nu~~,~~ R_{ijij}=R_{\mu\nu\lambda\rho}n_i^\mu n_j^\nu n_i^\lambda n_j^\rho~~~.
\end{equation}
The entropy of stationary black holes in such theories is 
\begin{equation}\label{5.24}
S^{BH}(G, c_i)={1 \over 4G}{\cal A}-\int_{\cal B}d^2x\sqrt{\gamma} \left(8\pi c_1 R+4\pi c_2 R_{ii}+8\pi c_3 R_{ijij}\right)~~~,
\end{equation}
where $\gamma=\det \gamma_{\alpha\beta}$, and $\gamma_{\alpha\beta}$ is defined in (\ref{2.20}). 

An alternative way to derive (\ref{5.24}) is to use the off-shell approach or the conical singularity method.
As was shown in  \cite{Fursaev:1995ef} integrals of powers of the curvature tensor can be well-defined on off-shell 
instantons with conical singularities in the linear approximation in the deficit angle. Such integrals are similar to (\ref{5.36}) 
and yield (\ref{5.24}) if one applies  (\ref{5.38}) for definition of the entropy.

\subsection{Generalized second law}\label{gen1}

By considering classical processes
with black holes one can conclude that the area of the horizon
never decreases, the observation which is reminiscent to the
second law.  Black hole must have an intrinsic entropy proportional
to the horizon area, otherwise processes like a gravitational collapse
would be at odds with the second law. The second law of thermodynamic in the presence of black holes can be written in the generalized form
\begin{equation}\label{5.25}
\delta S^{BH}+\delta S_m\geq 0~,
\end{equation}
which states that the sum of the Bekenstein-Hawking entropy of a black hole and the entropy of a surrounding  matter  
does not decrease in physical processes.

Generalized second law (\ref{5.25}) poses a number of serious questions  when it is applied to quantum matter around black holes.
The very definition of $S_m$, when the black hole horizon serves as a boundary of the system, requires clarification. Thermal entropy of a relativistic 
plasma is not well defined near $\cal H$ due to an infinite blue-shift of the energies of quanta in this region.
The classical part, $S^{BH}$, of the generalized entropy depends on the gravitational coupling $G$ which participates in the renormalization of the ultraviolet
divergences. Can $S_m$  in (\ref{5.25}) be considered as a quantum correction to the classical entropy $S^{BH}$? 

Although such questions may look technical, their resolution is a necessary step toward understanding profound conceptual issues 
brought in theoretical physics by black holes. Perturbative quantum gravity is a testbed where these questions can be dealt with by using conventional quantum field theory.

\section{Quantum black holes in thermal equilibrium}\label{QBH}
\subsection{The Hartle-Hawking-Israel state}

To proceed with the discussion of generalized second law (\ref{5.25}) in case of 
quantum fields around a black hole one needs to specify the quantum state of the system. If a black hole has an astrophysical mass  it evaporates due to the Hawking effect very slowly, as if being in a thermal equilibrium with the radiation. A real equilibrium state
can be realized for an eternal black hole placed inside a cavity with perfectly reflecting walls. A classical analogue of this system has been studied in \cite{York:1986it}. 

Consider, for simplicity, quantum fields 
on space-time of an eternal Schwarzschild black hole. This black hole has two space-like singularities, the future (black hole) singularity, the past (white) hole singularities, and two asymptotically flat, left and right regions, separated by the horizons, see Fig. \ref{fig2}.
The horizons ${\cal H}^\pm$ bifurcate at a two sphere $\cal B$. The structure of the Killing field of the Schwarzschild black hole is almost identical to that of Minkowsky space-time. This fact indicates that, for a black hole, there may be defined a quantum state which is 
a counterpart of the Minkowsky vacuum. Such a state does exist and is called the Hartle-Hawking-Israel state (HHI-state)
\cite{Hartle:1976tp},\cite{Israel:1976ur}. Stationary observers which move along integral lines of $\partial_t$ are analogous to the Rindler observers, and they  see the HHI state as a thermal bath
at the Hawking temperature. 

To come to the definition of this equilibrium state we use results of Sec. \ref{eucl}. The fact that classical gravity formulated 
on gravitational instantons has a thermodynamic form indicates also the importance of quantum theory on Riemannian (Euclidean) manifolds.
From now on we use notation ${\cal M}_E$ for these geometries.  Euclidean QFT has mathematical advantages, which can be explained
by using example of a free scalar field $\phi$ with equation
\begin{equation}\label{6.1}
P_E\phi=(-\nabla^2+m^2)\phi=0~~
\end{equation}
on ${\cal M}_E$. Operator $P_E$ is of a Laplace type. 
Since ${\cal M}_E$ is a curved manifold plane waves $e^{ikx}$ are not eigen-functions of $P_E$.
However one can act by $P_E$ on a plane wave to get
\begin{equation}\label{6.2}
P_Ee^{ik_\mu x^\mu}=\left[k_\mu k_\nu g_E^{\mu\nu}(x)+...\right]e^{ik_\mu x^\mu}~~.
\end{equation}
In mathematical applications, like the spectral theory, the important property of $P_E$ is that its leading symbol 
$\sigma_P(x,k)=k_\mu k_\nu g_E^{\mu\nu}(x)$
is not degenerate and is positive-definite on ${\cal M}_E$ (so $P_E$ are called elliptic operators). This behavior of the leading symbols 
is crucially different for operators on Euclidean and Lorentzian manifolds.  For elliptic operators spectral functions, which serve to define other ingredients of the quantum theory,
such as the effective action, can be introduced with mathematically meaningful prescriptions since large $k^2$ asymptotics are under control.
More on this topic can be found in monograph \cite{Fursaev:2011zz}. 

One of the key quantities which can be rigorously defined is the heat kernel $K(x,y|t)$ of $P_E$  which is the solution
to the following problem:
\begin{equation}\label{6.3}
(\partial_t+P_E(x))K(x,y|t)=0~~,~~K(x,y|t)_{t\to 0}=\delta(x,y)~~.
\end{equation}
Here $t$ is a positive parameter, $P_E(x)$  acts on argument  $x$ , $\delta(x,y)=\delta^{4}(x-y)/\sqrt{g}$, and the symmetry, $K(x,y|t)=K(y,x|t)$, is implied. Suppose that 
$P_E(x)$ does not have zero eigen-values. Then by using the heat kernel
one can define the Green function $G(x,y)$ of the operator
\begin{equation}\label{6.4}
P_E(x)G_E(x,y)=\delta(x,y)~~.
\end{equation}
One can check with help of (\ref{6.3}) that
\begin{equation}\label{6.5}
G_E(x,y)=\int_0^\infty dt~ K(x,y|t)~~.
\end{equation}

Hartle and Hawking \cite{Hartle:1976tp} used extension of $G_E(x,y)$, via the inverse Wick rotation to the corresponding Lorentzian 
space-time $\cal M$, to define the Green's function $G(x,y)$ on $\cal M$. It is a unique quantum state fixed in this way that is called the HHI-state.  

Below we briefly describe statistical-mechanical interpretation of Euclidean QFT's for quantum fields on stationary space-times.
Let $\cal M$ be a stationary space-time with the time-like Killing field $\zeta=\partial_t$. Let $g_{\mu\nu}$ be components
of metric of  $\cal M$ in coordinates $x^\mu(t)=(t,x^i)$. First suppose that this coordinate chart covers $\cal M$ globally, that is 
$\cal M$ has the structure $\Sigma \times R^1$ where $\Sigma$ are constant $t$ hypersurfaces. Consider decomposition (\ref{4.9}) of the field 
$\phi(x)$ on modes which are eigen-functions of the operator $\zeta=\partial_t$, see (\ref{4.11}). Since $\cal M$  is stationary one can define a finite-temperature 
sate of $\phi$ at temperature $T=\beta^{-1}$. The average of an operator $\hat{\cal O}$ in this state is: 
\begin{equation}\label{6.6}
\langle \hat{\cal O}\rangle_\beta=Z^{-1}(\beta)\mbox{Tr}(\hat{\cal O}e^{-\beta :\hat{H}:})~~,
\end{equation} 
where $Z(\beta)$ is partition function (\ref{5.30}) and the Hamiltonian $:\hat{H}:$  generates evolution along  $t$.
One is usually interested in a relation between the Wightman functions
\begin{equation}\label{6.7}
G_\beta^+(x,x')=\langle \hat{\phi}(x)\hat{\phi}^+(x')\rangle_\beta~~,~~G_\beta^-(x,y)=\langle \hat{\phi}^+(x') \hat{\phi}(x)\rangle_\beta~~
\end{equation} 
on $\cal M$  and the Green's function $G_E(x,y)$ on ${\cal M}_E$. It is assumed that $\cal M$ and ${\cal M}_E$ are connected via the Wick rotation 
$t\to -i\tau$. The inverse components of the Lorentzian and Euclidean metrics are: $g_E^{\tau\tau}=-g^{tt}$, $g_E^{\tau k}=ig^{tk}$, $g_E^{ik}=g^{ik}$.
In general, ${\cal M}_E$ is a complex manifold and $P_E(x)$ is not self-adjoint but it is still elliptic, which is enough to define with its help
corresponding spectral functions and the effective action. 

Let $x(t)$ be points on an integral line of $\partial_t$ on $\cal M$ and $\bar{x}(\tau)=x(-i\tau)$ be points on corresponding integral line 
of $\partial_\tau$ on ${\cal M}_E$.
One can define the two-point function 
\begin{equation}\label{6.8}
\tilde{G}_\beta(x(z),x'(0))=\theta(-\Im~z)G_\beta^+(x(z),x'(0))+\theta(\Im~z)G_\beta^-(x(z),x'(0))~~.
\end{equation}
where $z=t+i\tau$. It can be shown, see e.g.  \cite{Fursaev:2011zz} that (\ref{6.8}) is an analytic function of $z$ everywhere in the strip 
$-\beta < \Im z < \beta$ except the domains where the Wightman functions have singularities, that the periodicity property,
$\tilde{G}_\beta(x(z-i\beta),x'(0))=\tilde{G}_\beta(x(z),x'(0))$, holds, and that there is the fundamental relation,
\begin{equation}\label{6.9}
G_E(\bar{x}(\tau),x(0))=\tilde{G}_\beta(x(-i\tau),x'(0))~~,
\end{equation} 
between Euclidean and thermal Green's functions.

In case of black holes coordinates  $x^\mu(t)=(t,x^i)$ cover only a part of $\cal M$ located to the right from the horizon ${\cal H}^+\cup {\cal H}^-$, see Fig. \ref{fig2}. Constant time hypersurfaces $\Sigma$ intersect on $\cal B$.   The analysis \cite{Hartle:1976tp} shows that if the Eucliden Green's function in (\ref{6.9})  is defined on a black hole instanton, at $\beta=\beta_H=1/T_H$, 
the corresponding finite-temperature Green's function can be extended to the entire black hole space-time $\cal M$, beyond the
domain of stationary coordinates. The quantum state defined by such a Green's function is the Hartle-Hawking-Israel state.
(When applying this analysis to Kerr black holes one should note that 
$\zeta$ is time-like in a restricted domain and that non-diagonal components of ${\cal M}_E$ can be made real by going to imaginary values of the angular momentum, see Sec. \ref{eucl}.)  If $\beta$ is an arbitrary parameter there are some peculiarities in the behavior
of the Green's functions due to conical singularities discussed in next sections.

By the construction, a Killing observer sees the HHI state as a thermal bath 
at the Hawking temperature $T_H$. In the near-horizon approximation (\ref{5.8a}),(\ref{5.8b})
the single-particle excitations for such observers are defined by the set of left-moving and right-moving modes
\begin{equation}\label{6.10}
f^{(+)}_{L,\omega}(v)={1 \over \sqrt{4\pi \omega}}e^{-i\omega v}~~,~~f^{(+)}_{R,\omega}(u)={1 \over \sqrt{4\pi \omega}}
e^{-i\omega u}~~,
\end{equation}
which are eigen-functions of $\zeta=\partial_t$. In case of an eternal Schwarzschild black hole modes $f^{(+)}_{L,\omega}$
start from $\mathcal{I}^{-}$ and end on the black hole singularity, while $f^{(+)}_{R,\omega}$ start at the white hole singularity
and move to $\mathcal{I}^{+}$, see Fig. \ref{fig2} . 

One can also consider modes 
\begin{equation}\label{6.11}
\bar{f}^{(+)}_{L,\omega}(v)={1 \over \sqrt{4\pi \omega}}e^{-i\omega V(v)}~~,~~\bar{f}^{(+)}_{R,\omega}(u)={1 \over \sqrt{4\pi \omega}}e^{-i\omega U(u)}~~,
\end{equation}
where $V$ and $U$ are Kruskal coordinates (\ref{5.9}). The HHI state is the vacuum state for quasiparticles associated with modes (\ref{6.11}). 
To check this one can calculate
the number of particles related to (\ref{6.10}) in a vacuum state for (\ref{6.11}) by using 
(\ref{5.4}). The relevant non-trivial Bogoliubov coefficients between $R$-modes are
\begin{equation}\label{6.12}
d^R_{\sigma,\omega}=\langle f^{(+)}_{R, \sigma},\bar{f}^{(-)}_{R, \omega}\rangle=-{1 \over 2\pi}
\sqrt{\sigma \over \omega}\int^\infty_{-\infty} du ~e^{i\sigma u+i\omega U(u)}~~.
\end{equation}
As follows from (\ref{5.15}), the integral in the r.h.s. of (\ref{6.12}) is given by (\ref{5.16}). Thus  (\ref{6.12})  corresponds to 
the thermal distribution of quanta at the Hawking temperature, which is the property of the HHI state. The corresponding coefficient between the left modes has analogous expression. 

The HHI state can be defined on a global Cauchy surface where it has an important 
representation. Consider, as an example, the Carter-Penrose diagram of an eternal Schwarzschild black hole. A constant time
hypersurface $\Sigma_R \cup \Sigma_L$ which goes from the right to the left world is called the 
Einstein-Rosen bridge, see Fig. \ref{fig2}. The left and right parts of the bridge are identical and coincide with constant  time sections.
In  HHI wave function, in the configuration representation, depends on field variables ${\varphi_R}$, ${\varphi_L}$ set on 
$\Sigma_R$, $\Sigma_L$, correspondingly.
The wave function can be represented as transition amplitude in the Euclidean time $\tau$ between the left and right
worlds,
\begin{equation}\label{6.25}
\langle {\varphi_L}, {\varphi_R}|HHI \rangle=N^{-1/2}\langle {\varphi_L}|e^{-{\beta_H \over 2} :\hat{H}:}|{\varphi_R}\rangle=
\int [D\phi]e^{-I_E[{\varphi_L}, {\varphi_R}]}~~,
\end{equation}
where $N$ is a normalization factor, and the Euclidean action,
\begin{equation}\label{6.26}
I_E[{\varphi_L}, {\varphi_R}]=-\frac 12 \int_0^{\beta_H/2}d\tau\int d^3x~\phi^*P_E\phi~~,
\end{equation}
is defined with boundary conditions $\phi(\tau=0,x^i)=\varphi_R(x^i)$, $\phi(\tau=\beta_H/2,x^i)=\varphi_L(x^i)$.  A detailed discussion
of this representation, which is a natural generalization of a similar formula for the Minkowsky vacuum \cite{Israel:1976ur},  
can be found in \cite{Barvinsky:1994jca}.

\subsection{Effective action and renormalized stress-energy tensor}

Vacuum polarization in an external gravitational field $g_{\mu\nu}$
results a non-trivial average 
of the stress energy tensor of a quantum field,
$\langle \hat{T}_{\mu\nu}\rangle$, which appears in right-hand side of the
Einstein equations (\ref{i.1}) .  
The advantage of the Euclidean formulation of the theory is that $\langle \hat{T}_{\mu\nu}\rangle$ in the HHI state 
can be derived by variation of the effective action.

An introduction to the effective action approach in perturbative quantum gravity can be found in \cite{Buchbinder:1992rb}.
The effective action  of non-interacting quantum fields can be defined as 
the following functional on ${\cal M}_E$:
\begin{equation}\label{6.13}
\Gamma [g]= I [g,(G^B,\Lambda^B,c_i^B)] +W[g]~~.
\end{equation}
Here $I[g]$ is classical gravity action (\ref{5.22}) and $G^B,\Lambda^B,c_i^B$ are bare couplings.
The quantum part of the action is  
\begin{equation}\label{6.14}
W[g]={\eta \over 2} \ln \det P_E~~,
\end{equation}
where $\eta=+1$ or  $-1$ for fields with Bose or Fermi statistics, respectively. Expression (\ref{6.14}) is motivated by 
a formal path integral for free fields. It needs a further prescription to deal 
with ultraviolet divergences. Since $P_E$ is an elliptic operator one can introduce the heat trace (see details  
in \cite{Fursaev:2011zz})
\begin{equation}\label{6.15}
K(P_E;t)=\sum_{\lambda}e^{-t\lambda}~~,
\end{equation}
where the sum is taken over all eigenvalues $\lambda$ of $P_E$. If $\lambda$ are 
positive one can use the definition:
\begin{equation}\label{6.16}
\ln \det P_E=-\int_\delta^\infty {dt \over t}K(P_E;t)~~
\end{equation}
where $\delta>0$ is a proper cutoff parameter. An important property of (\ref{6.15}) is a short $t$ expansion (as $t\to+0$)
\begin{equation}\label{6.17}
K(P_E;t)\sim \sum_{p=0}^\infty t^{p-n \over 2} a_p(P_E)~~,
\end{equation}
where $n$ is the dimensionality of ${\cal M}_E$, and  $a_p(P_E)$ with odd $p$ appear if ${\cal M}_E$ has boundaries. 
The heat kernel (or DeWitt-Seeley) coefficients $a_p(P_E)$ determine the divergent part the effective action 
\begin{equation}\label{6.18}
W_{\tiny\mbox{div}}[g,\delta]=\eta \sum_{p=0}^{n-1}  { a_p(P_E) \over p-n}\delta^{p-n \over 2} +\eta a_n(P_E)\ln \delta~~,
\end{equation}
where we used (\ref{6.16}). A common prescription to eliminate the ultraviolet divergences
is to note that $a_p(P_E)$ with $p=2k$ are integrals of the $k$-th order polynomials in the curvature 
tensor. Thus, in four dimensions  (\ref{6.18}) has the same structure as the bare action $I [g, (G^B,\Lambda^B,c_i^B)]$.
The divergences are eliminated by redefinition of $G^B,\Lambda^B,c_i^B$,
\begin{equation}\label{6.19}
I [g, (G,\Lambda,c_i)]=I [g, (G^B,\Lambda^B,c_i^B)]+
W_{\tiny\mbox{div}}[g,\delta]~~~.
\end{equation} 
This yields effective action (\ref{6.13})  in the renormalized form:
\begin{equation}\label{6.20}
\Gamma [g]= I [g, (G,\Lambda,c_i)] +W_{\tiny\mbox{ren}}[g]~~,
\end{equation}
where renormalized, or the UV-finite part is 
\begin{equation}\label{6.21}
W_{\tiny\mbox{ren}}[g]=\lim_{\delta \to 0}(W[g,\delta] - W_{\tiny\mbox{div}}[g,\delta])~~.
\end{equation}
Formula (\ref{6.20}) can be used to derive the gravity equations which allow one to take into account the back-reaction to quantum fields,
\begin{equation}\label{6.22}
R_{\mu\nu}-\frac 12 g_{\mu\nu}R+...={8\pi G \over c^4}T^E_{\mu\nu} ~~,
\end{equation}
\begin{equation}\label{6.23}
T^{E~\mu\nu} [g]={2 \over \sqrt{\det |g|}}{\delta W_{\tiny\mbox{ren}}[g] \over \delta g_{\mu\nu}}~~.
\end{equation}
The dotes in  (\ref{6.22}) are quadratic in curvature terms.
To calculate first quantum correction to the metric tensor $g_{\mu\nu}$ it is enough
to consider $T^{E~\mu\nu} [g]$ on corresponding classical black hole instanton. 

Several remarks are in order.

i)  One can analytically continue (\ref{6.22}) from the Euclidean to the Lorentzian theory. If we are interested in first quantum
correction, the Lorenzian theory is just the gravity theory
sourced by the average of the stress-energy tensor in the Hartle-Hawking-Israel state. On black hole instanton
\begin{equation}\label{6.24}
T^E_{\mu\nu}[g]=i^q\langle HHI |\hat{T}_{\mu\nu} [g]|HHI \rangle~~,
\end{equation}
where $q$ is the number of temporal indexes and the factor $i^q$ is related to the Wick rotation. Given connection (\ref{6.9})
between the Euclidean and finite-temperature Green's functions  relation (\ref{6.24})  
can be proved when the quantum stress-energy tensor is computed with the help of a Green's function by using well-known
point-splitting procedure. (The point-splitting method is discussed in detail in \cite{Birrell:1982ix}.)

ii) By the construction, (\ref{6.24}) yields the stress-energy tensor in stationary coordinates outside the horizon. 
Since the Green's function
can be analytically extended to the entire space-time, so does (\ref{6.24}).

iii)  In addition to its transparent physical meaning the HHI state is distinguished by  mathematical properties when computations of quantum averages 
are reduced to finding various spectral functions of elliptic operators $P_E$. 

iv) There are different types of UV regularizations in the effective action, for eaxmple, the $\zeta$-function regularization
\cite{Hawking:1976ja},  
\cite{Elizalde:1994gf},\cite{Elizalde:1995hck}, the dimensional regularization and the Pauli-Villars regularization \cite{Birrell:1982ix}.  Up to
several finite counter-terms all of them yield the same  $W_{\tiny\mbox{ren}}$.

\section{Fluid dynamics in gravitational fields}\label{RHD}
\subsection{Finite-temperature QFT's and effective action}

The connection between the Euclidean and finite-temperature Green's functions suggests that there may 
exist an analogous relation between the Euclidean effective action $W$ and the free-energy of the quantum field. 
Let us start with stationary space-times  of the structure $\Sigma \times R^1$, 
where $\Sigma$ are constant time sections, and assume a time-like Killing field acts globally on $\cal M$.  

For simplicity we proceed with a free scalar field. Generalization to other fields is possible but requires additional
irrelevant details. Wave equation (\ref{4.1}) on a stationary space-time can be written as
\begin{equation}\label{7.1}
P(\partial_t,\partial_i)\phi=0~~~.
\end{equation}
The relation between Lorentzian and Euclidean operators is  $P_E(\partial_\tau,\partial_i)=P(i\partial_\tau,\partial_i)$.
To define statistical-mechanical quantities one needs the single-particle energies $\omega_i$ introduced in (\ref{4.11}).
Their spectrum is determined by the problem
\begin{equation}\label{7.2}
(P_0\omega^2+P_1\omega +P_2)\phi_\omega=0~~,
\end{equation}
which follows after substitution $\phi(t,x^i)=e^{-i\omega t}\phi_\omega (x^i)$ to (\ref{7.1}). 
Here $P_k$ are $k$-th order partial differential operators. Eq. (\ref{7.2}) is called non-linear spectral problem.

With help of the single-particle spectrum the free energy of the considered system can be written as
\begin{equation}\label{7.3}
F(\beta)= -\beta^{-1}\ln Z(\beta)=\beta^{-1}\sum_{i} \ln\left(1-e^{-\beta \omega_i}\right)~~,
\end{equation}
where $Z(\beta)$ is  given by (\ref{5.30}). The summation goes over all  single-particle
energies $\omega_i^{\pm}$. For continuous spectra the sum is replaced with corresponding integrals.

One can formally define the vacuum energy 
$E_0=\frac 12\sum_{i}\omega_i$, where the series diverges and should be considered in the framework of some regularization 
prescription. Let  $E_{0,\tiny\mbox{ren}}$ be a finite (renormalized) part of $E_0$ left after subtracting divergent terms.
Then there is the relation between the free energy and Euclidean action 
(\ref{6.14}):
\begin{equation}\label{7.4}
W_{\tiny\mbox{ren}}[g]=\beta(F(\beta)+E_{0,\tiny\mbox{ren}})~~.
\end{equation}  
The left and right parts of (\ref{7.4}) coincide up to finite counterterms. Derivation of (\ref{7.4}) can be found, e.g.
in \cite{Fursaev:2011zz}.

On the considered class of geometries $F(\beta)$ is finite at large $\omega_i$. As follows from (\ref{7.4}) the UV divergent part
of the effective action appears from the divergences of the vacuum energy.  As we see in Sec. \ref{gene}, the situation changes when the Killing field $\zeta$
has the horizon.

\subsection{High-temperature asymptotic}\label{HT}

A Killing observer with 4-velocity along $\zeta=\partial_t$ measures the so called local Tolman temperature
\begin{equation}\label{7.5}
T(x)={T_0 \over \sqrt{-\zeta^2}}~~,
\end{equation}  
where $T_0 =\beta^{-1}$. In asymptotically flat space-times $T_0$ is a
temperature measured by observers at infinity. Factor $\sqrt{-\zeta^2}$ relates coordinate time $t$ to the proper time of the observer.
The Tolman temperature increases if position of the observer is taken closer to the black hole horizon. One says that 
temperature is blueshifted near $\cal H$. 

Therefore in the near-horizon region the system is effectively at a high temperature regime. Interestingly, 
high-temperature limit allows an analytic form of the free energy which, in general, can be written in terms 
of characteristics of the Killing frame of reference, discussed in Sec. \ref{def}. For example, in the absence of boundaries
the leading terms look as
$$
F(\beta)\simeq -\int \sqrt{-g}d^3x \left[b_1T^4(x)+\right.
$$
\begin{equation}\label{7.6}
\left. T^2(x)(b_2R+b_3 m^2+b_4~ \Omega^2(x)
+b_5~w_\mu w^\mu+b_6~\nabla w)+O(\ln\beta)\right]~~.
\end{equation}  
Here $w_\mu$ is 4-acceleration, $\Omega=\frac 12(A_{\mu\nu}A^{\mu\nu})^{1/2}$ is the
absolute value of the local angular velocity, $A_{\mu\nu}$ is the rotation tensor, see definitions (\ref{2.13}). The structure of (\ref{7.6}) 
is determined by using canonical mass dimensions of all quantities, $b_k$ are  
numerical coefficients which depend on the considered model and 
require computations.  For a real scalar field with equation $(-\nabla^2+V)\phi=0$ one 
finds \cite{Fursaev:2001yu}
\begin{equation}\label{7.7}
F(\beta)\simeq -\int d^3x \sqrt{-g}\left[{\pi^2 \over 90}  T^4
+ {1 \over 24} T^2\left(\frac 16 R-V-\frac 23 \Omega^2
\right) +O(\ln\beta)\right]~~.
\end{equation}
According with (\ref{7.4}) one can split the stress-energy tensor as
\begin{equation}\label{7.8} 
\langle HHI |~\hat{T}_{\mu\nu}~ | HHI \rangle
=\langle ~\hat{T}_{\mu\nu}~\rangle_\beta+\langle ~\hat{T}_{\mu\nu}~ \rangle_{\mbox{vac}}~~.
\end{equation}
Thermal part $\langle T_{\mu\nu}\rangle_\beta$ is determined by the free energy $F(\beta)$,
while the vacuum part $\langle T_{\mu\nu}\rangle_{\mbox{vac}}$ follows from the vacuum energy.
Variations of (\ref{7.7}) over the metric yield the thermal part at high temperatures.
For example, its expression in the case of conformal coupling, $V=R/6$, is \cite{Fursaev:2001yu}
$$
\langle T_{\mu\nu}\rangle_\beta\sim(g_{\mu\nu}+4u_\mu u_\nu) \left({\pi^2 \over 90}
T^4-{1 \over 36} T^2\Omega^2\right)
$$
\begin{equation}\label{7.9} 
+{T^2 \over 36}\left(2A_{\nu\rho}A_{\mu}^{~~\rho}-u_\nu
A_{\mu\lambda} w^\lambda-u_\mu A_{\nu\lambda}w^\lambda \right.
\left. +u_\nu A_{\mu\lambda}^{~~~;\lambda}+u_\mu
A_{\nu\lambda}^{~~~;\lambda} \right)+O(\ln \beta).
\end{equation}
One can check that $\nabla^\mu\langle T_{\mu\nu}\rangle_\beta=0$, $g^{\mu\nu} \langle T_{\mu\nu}\rangle_\beta=0$.
High-temperature asymptotics in static space-times have been studied in pioneering papers 
\cite{Dowker:1978md},\cite{Dowker:1988jw}. Extension to stationary geometries, which requires one to deal with 
non-linear spectral problems  (\ref{7.2}) has been suggested in \cite{Fursaev:2001yu}, see \cite{Fursaev:2011zz} for more details.

Asymptotics (\ref{7.6}), (\ref{7.7}) are of increasing interest in modern studies
of equilibrium distributions in finite-temperature quantum field theories with rotation and acceleration in flat space-times
\cite{Becattini:2020qol}, where $A_{\mu\nu}$ is interpreted as a thermal vorticity. The interest is related to properties of quark-gluon plasma
in heavy ion collisions.

\section{Generalized thermodynamics of quantum black holes}\label{GTD}
\subsection{Generalized free energy}

The HHI state describes an eternal black hole in thermal equilibrium with its radiation. The system, a black hole
and quantum fields in the HHI state, can be called a quantum black hole. The split of 
this system on classical and quantum components is rather conditional. The classical black hole geometry backreacts to quantum matter, 
while bare classical couplings participate in renormalization of ultraviolet divergences.

Black hole thermodynamics poses a number of fundamental questions to quantum gravity mentioned in previous Sections. 
As a first step in resolving the existing paradoxes, it is reasonable to understand thermodynamics of quantum black holes in HHI state.
We describe here some steps how it can be done by using advantages of the Euclidean field theory. For simplicity, we consider
non-rotating black holes. Extension of the arguments below to black holes with angular momentum is possible but it does not
carry principle issues.

The key assumption we adopt is that the generalized free energy of a quantum black hole is determined by the effective action
$\Gamma[g]$, see (\ref{6.20}). The background metric, $\bar{g}$, is a stationary point of  $\Gamma[g]$
under certain boundary conditions, which can be defined  as in classical case \cite{York:1986it}. 
One also requires that $\bar{g}$ is a black-hole-instanton-like solution with a global Euclidean Killing vector field $\partial_\tau$,  $\tau$ 
being a periodic
coordinate with a period $\beta$. We interpret $\beta$ as an inverse temperature.

Variations of $\Gamma[g]$ lead to the Einstein equations (\ref{6.22}) sourced by the stress-energy tensor of the quantum matter.
The functional $W[g]$ is a complicated non-local functional which is known in some approximations, in the second order curvature approximation \cite{Barvinsky:1985an}, for example. Quantum corrected Schwarzschild solutions  in the framework of these approximations have been
discussed in \cite{Calmet:2021lny},\cite{Xiao:2021zly}.

The generalized free-energy of a quantum black hole is defined as follows:
\begin{equation}\label{8.1} 
F_{\tiny\mbox{gen}}(\beta)=\beta^{-1}\Gamma[\bar{g}(\beta)]~~.
\end{equation}
There are some subtle issues related to normalization of $\Gamma[\bar{g}(\beta)]$ in (\ref{8.1}) by analogy with the 
Gibbons-Hawking subtraction procedure \cite{Gibbons:1976ue}. We do not imply any subtraction in the quantum part of 
the effective action before we find its statistical-mechanical interpretation.
Eq. (\ref{8.1}) allows one to introduce generalized energy and entropy of the black hole by using standard definitions of statistical 
physics
\begin{equation}\label{8.2} 
E_{\tiny\mbox{gen}}(\beta)=\partial_\beta(\beta F_{\tiny\mbox{gen}}(\beta))~~,~~S_{\tiny\mbox{gen}}(\beta)=\beta^2
\partial_\beta(F_{\tiny\mbox{gen}}(\beta))~~.
\end{equation}
By virtue of (\ref{8.2}) parameters of two black holes with slightly different 
temperatures are related by the  first law
\begin{equation}\label{8.3} 
\delta E_{\tiny\mbox{gen}}(\beta)=\beta^{-1} \delta S_{\tiny\mbox{gen}}(\beta)~~,
\end{equation}
which is the generalization of (\ref{5.20}) for $\Omega_H=0$.  

The generalized free energy and entropy can be found in some cases, for example for a Schwarzschild black hole 
with massless quantum fields \cite{Fursaev:1994te},\cite{Solodukhin:1994yz}.  This can be done by using scaling properties 
of $W_{\tiny\mbox{ren}}(\beta)$, and leads to logarithmic corrections to the Bekenstein-Hawking entropy. 
The result can be  generalized to include higher loops  \cite{Solodukhin:2019xwx}. Corrections for other types of 
black holes are discussed in \cite{Reall:2019sah},\cite{Solodukhin:1994st}.

To proceed with (\ref{8.2})  we note that derivative of the effective action 
over $\beta$ can be taken in two steps,
\begin{equation}\label{8.4} 
\partial_\beta \Gamma[\bar{g}]=\partial_\beta \Gamma[\bar{g}]_{\bar{g}}+\int d^4x\sqrt{\bar{g}}~{\delta \Gamma[\bar{g}] \over \delta g_{\mu\nu}(x)}~
\partial_\beta \bar{g}_{\mu\nu}(x)+B~~.
\end{equation}
The first term, $\partial_\beta \Gamma[\bar{g}]_{\bar{g}}$, is the derivative over $\beta$ when the bulk metric is fixed. This derivative changes 
the periodicity  of the instanton $\bar{g}$ and results in conical singularities at the Euclidean horizon $\cal B$. The second and third terms in 
the r.h.s. of (\ref{8.4}) come out when one differentiates the metric but keeps the periodicity fixed. The third term, $B$, appears from the "inner boundary" $\cal B$. This term is related to singularities on
$\cal B$. The "inner boundary" $\cal B$ in the off-shell approach should be taken into account 
when integrating by parts to get rid of derivatives of $\delta \bar{g}$.  Since there are no singularities on the physical boundary 
$\partial {\cal M}_E$ we assume that corresponding boundary terms in (\ref{8.4}) are excluded by appropriate boundary conditions. 

Since $\bar{g}$ is a stationary point of  $\Gamma[g]$ the second 
term in the r.h.s of  (\ref{8.4}) disappears, and one gets
\begin{equation}\label{8.5} 
E_{\tiny\mbox{gen}}(\beta)=\partial_\beta \Gamma[\bar{g}]_{\bar{g}}+B~~.
\end{equation}
\begin{equation}\label{8.6} 
S_{\tiny\mbox{gen}}(\beta)=(\beta \partial_\beta-1) \Gamma[\bar{g}]_{\bar{g}}+\beta B~~.
\end{equation}
It follows from (\ref{6.20}), properties of the classical Euclidean action (\ref{5.34}) and (\ref{8.2}) that
\begin{equation}\label{8.7} 
E_{\tiny\mbox{gen}}(\beta)=\bar{M}_H(\beta)+E_q(\beta)+B~~,
~~
S_{\tiny\mbox{gen}}(\beta)=\bar{S}^{BH}(\beta)+S_q(\beta)+\beta B~~.
\end{equation}
Here $\bar{M}_H(\beta)$ and $\bar{S}^{BH}(\beta)$ are the mass and Bekenstein-Hawking entropy (for simplicity we assume that renormalized couplings $c_i=0$ in (\ref{6.20})) computed by applying classical formula to the quantum corrected
instanton $\bar{g}$.  We recall that $\beta^{-1}$ is identified with the Hawking temperature. Other quantum corrections in 
(\ref{8.7}) are defined as
\begin{equation}\label{8.8} 
E_q(\beta)=\partial_\beta W_{\tiny\mbox{ren}}(\beta)~~,~~S_q(\beta)=\left(\beta \partial_\beta-1\right)W_{\tiny\mbox{ren}}(\beta)~~.
\end{equation}
Since we are interested in first order quantum corrections, it is enough to take $W_{\tiny\mbox{ren}}(\beta)$ 
on the classical instanton. 

\subsection{Energy of a quantum black hole}  

Consider first classical black holes. The total mass $M$ of a black hole, as measured at infinity, is defined by the differential mass formula \cite{Bardeen:1973gs} :
\begin{equation}\label{8.9} 
M=M_H+E~~,
\end{equation}
\begin{equation}\label{8.10} 
E=-\int_{\Sigma} d^3x \sqrt{-g}~T_0^0~~,
\end{equation}
where $M_H$ is the mass measured at the horizon, and $E$ is the energy of matter outside the horizon. 
The covariant stress-energy tensor of matter $T^\mu_\nu$ for field theories is determined by the variation of the corresponding action over the metric. The integral
in (\ref{8.10}) 
is defined on a constant time slice $\Sigma$ outside the black hole.

Equation (\ref{8.9}) suggests the following  natural form of the generalized energy of a quantum black hole:
\begin{equation}\label{8.11} 
E_{\tiny\mbox{gen}}(\beta)=\bar{M}_H(\beta)+E(\beta)~~,
\end{equation}
\begin{equation}\label{8.12} 
E(\beta)=-\int_{\Sigma} d^3x \sqrt{-g}~\langle HHI |~\hat{T}_0^0~ | HHI \rangle~~.
\end{equation}
$\bar{M}_H$ is identified with the mass at the horzon, while $E$ is replaced with the expectation value of the 
corresponding energy operator in the HHI state.

Consider now quantum correction $E_q$ in  (\ref{8.7}).  This correction is defined in (\ref{8.8}).
It follows from (\ref{7.4}) and thermal properties 
of the HHI state that
\begin{equation}\label{8.13} 
E_q(\beta)=\langle :\hat{H}:\rangle_\beta+E_{0,\tiny\mbox{ren}}=\langle HHI |~ \hat{H}~ | HHI \rangle~~,
\end{equation}
The r.h.s. of (\ref{8.13}) is the renormalized expectation value of the canonical energy operator $\hat{H}$ without
any normal ordering.

Definition of energy by (\ref{8.10}) in terms of the metric stress-energy tensor and definition of energy
as the Hamiltonian may differ by a total derivative which results in a non-vanishing term
when $\Sigma$ has $\cal B$ as
an internal boundary, where the Killing field bifurcate. One can show that \cite{Frolov:1997up},\cite{Fursaev:1998hr}
\begin{equation}\label{8.14} 
E=H-\beta Q~~.
\end{equation}
Here $\beta$ is inverse Hawking temperature determined by the surface gravity of $\zeta$, and
$Q$ is an integral over $\cal B$. The quantity $Q$ is always non-trivial when there are non-minimal couplings 
between dynamical fields and the background curvature.
It can be demonstrated  
for black holes in a general classical theory of gravity arising from a diffeomorphism invariant Lagrangian that \cite{Fursaev:1998hr}:

i) $E$ appears in the first law of black hole mechanics and plays the same role as the energy in the differential mass formula
(\ref{8.9});

ii) $H$ is the canonical energy which generates evolution along the Killing time;

iii) $Q$ in (\ref{8.11}) is a Noether charge analogous to Wald's charge;

iv)  the entropy of a black hole in the presence of non-minimal couplings is
\begin{equation}\label{8.15} 
S=S^{BH}-Q~~.
\end{equation}
To give a typical illustration of non-minimal 
coupling   consider the scalar theory
\begin{equation}\label{8.16}
I[\phi, g]=-\frac 12 \int d^4x \sqrt{-g} \left[(\nabla \phi)^2+
\xi R\phi^2+m^2\phi^2\right]~~.
\end{equation}
A direct check yields \cite{Frolov:1997up}
\begin{equation}\label{8.17}
Q=2\pi  \xi \int_{\cal B} d^2y~ \phi^2(y)~~.
\end{equation}
Other examples of non-minimal couplings with non-trivial charges are present for fields with integer spins. 

The quantum version of (\ref{8.14}) should be the following formula for the average 
energy (\ref{8.12}):
\begin{equation}\label{8.18} 
E(\beta)=\langle HHI |~ \hat{H}~ | HHI \rangle-\beta \langle HHI |~ \hat{Q}~ | HHI \rangle ~~.
\end{equation}
By comparing (\ref{8.11}) with (\ref{8.7}) one concludes that 
\begin{equation}\label{8.19} 
B=-\beta \langle  \hat{Q} \rangle ~~,
\end{equation}
where $\langle \hat{Q} \rangle=\langle HHI |~ \hat{Q}~ | HHI \rangle$.  As can be seen from (\ref{8.17}) 
the expectation value $\langle  \hat{Q} \rangle$ is determined by two-point correlators, like $\langle  \hat{\phi}(x) \hat{\phi}(x') \rangle$
in the limit $x\to x'$. Since such limits results in UV divergent terms, it is assumed that  $\langle  \hat{Q} \rangle$ is a renormalized quantity.

The presented arguments which lead to (\ref{8.19}) cannot be considered as a rigorous proof but they pass non-trivial 
checks which we consider in the next Section.

\subsection{Generalized black hole entropy and its interpretations}\label{gene} 

As follows from (\ref{8.7}), (\ref{8.19})  the generalized entropy of a quantum black hole is the sum of three terms, 
\begin{equation}\label{8.20} 
S_{\tiny\mbox{gen}}(\beta)=\bar{S}^{BH}(\beta)+S_q(\beta)-\langle  \hat{Q} \rangle~~.
\end{equation}
The quantum correction to black hole entropy includes contribution from non-minimal couplings $\langle  \hat{Q} \rangle$.
This fact has been first
pointed out in \cite{Solodukhin:1995ak}. The origin of $\langle  \hat{Q} \rangle$ in (\ref{8.20}) can be traced to 
the delta-function-like term in the scalar curvature on $\cal B$, which appears in the off-shell approach according to Eq. (\ref{5.36}). 
Let us emphasize that $\bar{S}^{BH}(\beta)$ in (\ref{8.20}) depends on the renormalized gravitational couplings, UV divergences 
in $S_q(\beta)$ and $\langle \hat{Q}\rangle$ are subtracted. 

Consistency of (\ref{8.20}) implies that divergences 
in $S_q(\beta)$ and $\langle \hat{Q}\rangle$ are absorbed in the course of renormalization of the bare couplings in $\bar{S}^{BH}(\beta)$.
To demonstrate this in the one-loop approximation we focus on properties and interpretation of $S_q(\beta)$.
To derive  $S_q(\beta)$ by using (\ref{8.8}) the one-loop effective action 
$W_{\tiny\mbox{ren}}(\beta)$ has to be considered off-shell, at $\beta$ 
slightly different from the Hawking temperature. We denote the corresponding background
manifold ${\cal M}_\beta$. Near the Euclidean horizon ${\cal M}_\beta$ has the structure ${\cal C}_\beta \times {\cal B}$, where
${\cal C}_\beta$ is a two-dimensional cone with metric
\begin{equation}\label{8.21} 
dl^2=\kappa^2 \rho^2d\tau^2 +d\rho^2~~,~~0\leq \tau <\beta~~.
\end{equation}
${\cal C}_\beta=R^2$, if $\beta=2\pi/\kappa$. The nice property of elliptic operators that their spectral functions
are well-defined on manifolds with conical singularities. 

Conical singularities modify results known for regular manifolds.
In particular, the heat coefficients of the heat trace (\ref{6.17}) on ${\cal M}_\beta$ with $p$ even look as
\begin{equation}\label{8.22}
a_p(P_E)=A_p+B_p~~,
\end{equation}
where $A_p$ are defined by standard expressions on regular domain of ${\cal M}_\beta$. Additional terms
$B_p$, $p=2k$, are some curvature polynomials on $\cal B$ which have non-trivial dependence on $\beta$. For example \cite{Fursaev:1994in},
\begin{equation}\label{8.22}
B_2(\beta)=f(\beta){\cal A}~~,~~B_4(\beta)=\int_{\cal B} \sqrt{\gamma}~d^2y~(g_1(\beta) R+g_1(\beta)R_{ii}+g_3(\beta)R_{ijij})~~,
\end{equation}
where $f(\beta)$, $g_k(\beta)$ are some functions analytic at $\beta=2\pi/ k$,
curvatures $R_{ii}$, $R_{ijij}$ are defined by (\ref{5.23}).
Definition (\ref{8.8}) implies that there must exist a divergent part 
of $S_q$ determined by the divergent part of the effective action. In four dimensions, $n=4$,
\begin{equation}\label{8.23}
S_{q,\tiny\mbox{div}}(\delta)=(\beta \partial_\beta-1)W_{\tiny\mbox{div}}\simeq 
s_1\delta +s_2\ln \delta+O(\delta^{-1})~~,
\end{equation}
\begin{equation}\label{8.24}
s_1=-{\eta  \over 2} (\beta \partial_\beta-1)B_{2}|_{\beta=\beta_H}~~,~~s_2=\eta(\beta \partial_\beta-1)B_{4}|_{\beta=\beta_H}~~
\end{equation}
To get (\ref{8.23}) we used (\ref{6.18}). 

Note that the leading divergence of $S_q$ is connected with $B_2$, and it is proportional to the horizon
area $\cal A$. This divergence behaves as the Bekenstein-Hawking entropy and, as has bee suggested in
\cite{Susskind:1994sm}, \cite{Callan:1994py}, it can be removed
by the standard renormalization of the Newton constant. 

Consider a standard renormalization procedure in the one-loop effective action.
To remove the divergences the bare gravity action should be taken in form (\ref{5.22}) with bare couplings $G^B,\Lambda^B,c_i^B$.
The relation between bare and renormalized couplings $G,\Lambda,c_i$ is (\ref{6.19}). Given (\ref{6.19}) 
one can show that the following
formula holds  \cite{Fursaev:1994ea},\cite{Solodukhin:1995ak},\cite{Frolov:1998vs}, \cite{Solodukhin:2011gn} :
\begin{equation}\label{8.25}
S^{BH}(G, c_i)=S^{BH}(G^B, c^B_i)+S_{q,\tiny\mbox{div}}(\delta)-\langle  \hat{Q}(\delta)\rangle_{\tiny\mbox{div}}~~~.
\end{equation}
The result holds for spins 0, 1/2 and 1. Formula (\ref{8.25}) is a non-trivial check which supports (\ref{8.20}).

One of the interpretations of $S_q$ is related to quantum entanglement in the HHI state since states on $\Sigma_L$ part
of the Einstein-Rosen bridge are not accessible for observers in the right part of the eternal black hole, see Fig. \ref{fig2}.
Consider a quantum system in a state $|\psi\rangle$. Suppose some part  $A$ of the system cannot be measured. 
$A$ may be, for example,  a spatial region. Let $B$ be a supplement of $A$.
Averages of operators located in $B$ can be written as 
\begin{equation}\label{8.26}
\langle \psi |\hat{\cal O}|\psi\rangle=\mbox{Tr}_{B}~(\hat{\rho}\hat{\cal O})~~,
\end{equation}
\begin{equation}\label{8.27}
\hat{\rho}=\mbox{Tr}_{A}~|\psi\rangle \langle \psi |~~.
\end{equation}
The density matrix $\hat{\rho}$  allows different measures of the information loss about states located in $A$. For example, one
can define
the Renyi entropy
\begin{equation}\label{8.28}
S^{(\alpha)}={\mbox{Tr}_{B}~\hat{\rho}^\alpha \over 1-\alpha}~~,~~\alpha>0~~,~~\alpha\neq 1~~,
\end{equation}
and the entropy of entanglement 
\begin{equation}\label{8.29}
S_{\tiny\mbox{ent}}=\lim_{\alpha\to 1}S^{(\alpha)}=-\mbox{Tr}_{B}~\hat{\rho}\ln \hat{\rho}~~.
\end{equation}
A review of quantum entanglement and entanglement measures can be found in \cite{Horodecki:2009zz}. Entanglement in quantum
field theories is discussed, e.g. in \cite{Calabrese:2004eu}.

In the considered case of an eternal black hole $A$ and $B$ are two sides of the Einstein-Rosen bridge separated by the 
bifurcation surface $\cal B$, say, 
$B=\Sigma_R$ and $A=\Sigma_L$. To calculate the entanglement entropy we 
use representation (\ref{6.25}) for the HHI state. If one assumes that $\langle HHI|HHI \rangle=1$,  the normalization
factor in (\ref{6.25}) is $N=\mbox{Tr}~e^{-\beta_H :\hat{H}:}=Z(\beta_H)$. This yields for natural parameters $n$
\begin{equation}\label{8.30}
\mbox{Tr}_{B}~\hat{\rho}^n={Z(n\beta_H) \over Z^n(\beta_H)}=\exp\left(-W(n\beta_H)+nW(\beta_H)\right)~~,
\end{equation}
where $W(\beta)$ should be considered as a regularized action.
To get (\ref{8.30}) we used (\ref{7.4}) and noticed mutual cancellation of the vacuum energies in $W(n\beta_H)$ and in $nW(\beta_H)$.  Since $W_{\tiny\mbox{ren}}(\beta)$ can be defined at arbitrary $\beta$
one can replace $n$ in the r.h.s. of (\ref{8.30}) with a parameter $\alpha>0$ and use (\ref{8.28}), (\ref{8.29}) to see that
\begin{equation}\label{8.31}
S_{\tiny\mbox{ent}}=S_q~~.
\end{equation}
That is the entanglement entropy in HHI state is given by formula (\ref{8.8}) and is a part of the 
generalized entropy (\ref{8.20}).

The fact that  $S_{\tiny\mbox{ent}}$ is proportional
to the horizon area and can be related to the Bekenstein-Hawking
entropy was first pointed out in \cite{Bombelli:1986rw}, \cite{Srednicki:1993im}, \cite{Frolov:1993ym}.
Since computations of the entanglement entropy  based on (\ref{8.8}) are reduced to finding effective actions on manifolds
with conical singularities $S_{\tiny\mbox{ent}}$ was also called a geometric entropy. Pioneering studies
of $S_{\tiny\mbox{ent}}$ can be found in \cite{Larsen:1994yt},\cite{Kabat:1994vj},\cite{Kabat:1995eq},\cite{Larsen:1995ax}.
A separate interesting topic is the entanglement entropy of gauge fields which is discussed in a number of publications, see, e.g. \cite{Casini:2019nmu}, \cite{Donnelly:2014fua}. A comprehensive review of entanglement entropy of black holes is \cite{Solodukhin:2011gn}.

One can also come to (\ref{8.31}) with the help of (\ref{6.25}) by noticing 
that the reduced density matrix for the HHI state is thermal, $\hat{\rho}=e^{-\beta_H :\hat{H}:}/Z(\beta_H)$. 
This means that the entanglement
entropy coincides with the entropy of the thermal
atmosphere of a black hole. The first attempts to relate
the Bekenstein-Hawking entropy to the thermal atmosphere
have been made in \cite{Zurek:1985gd},\cite{tHooft:1984kcu}. The thermal entropy can be derived from 
high-temperature asymptotic (\ref{7.6}) and it diverges due to the infinite blue-shift of the temperature at the horizon, see
(\ref{7.5}). To avoid divergences a narrow region near the horizon should 
be excluded when integrating in  (\ref{7.6}). If the cutoff is made at a proper distance equal the Planck length the thermal entropy is 
of the order of the Bekenstein-Hawking entropy \cite{tHooft:1984kcu}. The divergences in the thermal entropy can be 
eliminated by using other regularizations, for instance, the Pauli-Villars (PV) regularization \cite{Demers:1995dq}. Equivalence  between divergences of thermal entropy and entanglement entropy in different regularizations
is discussed in \cite{Frolov:1998vs}.

To simplify the presentation in last sections we restricted the discussion by static black holes.
The analysis can be extended to rotating black holes, see \cite{Frolov:1999gy} and references therein.

\section{Concluding remarks}\label{further}  

The aim of this Chapter was an introduction to generalized thermodynamics of quantum black holes and related notions
which appeared in last decades.  The key conclusion is that the Bekenstein-Hawking entropy is a part of
the generalized entropy (\ref{8.20}) of a black hole and it is deeply related to quantum effects via the renormalization of the ultraviolet
divergences. 

Renormalization requires the bare entropy which does not have any statistical meaning
in the perturbative quantum gravity. The problem of the bare entropy in  (\ref{8.25}) can be resolved
\cite{Jacobson:1994iw}
if the Einstein gravity is entirely induced by quantum effects, so that (\ref{8.20})
becomes $S_{\tiny\mbox{gen}}=S_q-\langle  \hat{Q} \rangle$.
This mechanism can be checked at a one-loop order in QFT models
where the leading ultraviolet divergences are canceled out \cite{Frolov:1996aj}, \cite{Frolov:1997up}.
The idea of induced gravity was formulated long ago
\cite{Sakharov:1967pk}, \cite {Sakharov:1975zg}, see 
\cite{Visser:2002ew} , for a review, but it does not contradict to a modern perspective. From the point of view of the open string theory
black hole entropy can be considered as a loop effect \cite{Hawking:2000da}, in full
analogy with its origin in induced gravity.  There still remains the problem of statistical interpretation of the "contact" term
$\langle  \hat{Q} \rangle$ in the entropy, whose presence is unavoidable to compensate divergences in $S_q$.

We have not discussed alternative interpretations of the black hole entropy which go beyond the perturbative 
quantum gravity.  A number of promising ideas to statistical explanation of the Bekenstein-Hawking formula
are reviewed in \cite{Carlip:2014pma}. Among them is a remarkable calculation of the entropy of extremal  \cite{Strominger:1996sh} and near-extremal black holes in string theory, more on that can be found in \cite{Peet:1997es}.  

There is a mounting evidence that the Bekenstein-Hawking entropy formula
can be applied for systems which are not black holes. For instance, in higher dimensional anti-de Sitter
gravities it allows successful holographic description \cite{Nishioka:2009un}, \cite{Ryu:2006bv} 
of the entanglement entropy in dual conformal field theories. In a similar way, ${\cal A}/(4G)$ can be interpreted as
as an entanglement entropy in low-energy limit of quantum gravity for regions restricted by
minimal surfaces of the area $\cal A$ \cite{Fursaev:2001yu},\cite{Lewkowycz:2013nqa}.

\section{Acknowledgements}

The author is grateful to I. Pirozhenko for the help with preparation of the figures and to G. Prokhorov for valuable discussions.

\end{document}